%% file: double_0th.tex
\documentclass[journal]{IEEEtran}
\ifCLASSINFOpdf
\else
\fi
\ifCLASSOPTIONcompsoc
 \usepackage[caption=false,font=normalsize,labelfont=sf,textfont=sf]{subfig}
\else
 \usepackage[caption=false,font=footnotesize]{subfig}
\fi

\usepackage{graphicx}
\usepackage{amsmath}
\usepackage{mathtools}
\usepackage{epsfig}
\usepackage{float}
\usepackage{lipsum}
\usepackage{stfloats}
\usepackage{array}
\usepackage{amssymb}
\usepackage{cite}
\usepackage{url}
\usepackage{algorithm2e}
\usepackage{algorithmic}
\usepackage{multirow}
\usepackage{fancyh}
\usepackage{upgreek}
\usepackage{dsfont}
\usepackage{gensymb}

\usepackage{multicol, blindtext}
\usepackage{mathtools, cuted}

\usepackage{color}
\usepackage{bm}
\usepackage{booktabs}
\usepackage{tablefootnote}
\usepackage{threeparttable}

\DeclareFontFamily{U}{mathc}{}
\DeclareFontShape{U}{mathc}{m}{it}%
{<->s*[1.03] mathc10}{}
\DeclareMathAlphabet{\mathscr}{U}{mathc}{m}{it}

\normalsize
\input{input.tex}
\graphicspath{ {./Images/} }

\begin{document}

% \title{Wideband Hybrid Beamforming\\for Satellite Communications
\title{Hybrid Arrays: How Many RF Chains Are Required to Prevent Beam Squint?
%Beam Squint in Hybrid Arrays: When is True Time Delay Required?
\thanks{H. Do and N. Lee are with the School of Electrical Engineering, Korea University, 02841
Seoul, South Korea (e-mail: \{doheedong, namyoon\}@korea.ac.kr). Their work is supported in part by the Institute of Information \& Communications Technology Planning \& Evaluation (IITP) under Grant 2021-0-00161 and in part by the National Research Foundation of Korea (NRF) under Grant 2023-00208552, both funded by the Korea Government (MSIT). A. Lozano is with Univ. Pompeu Fabra, 08018 Barcelona (e-mail: angel.lozano@upf.edu). His work is supported by the Maria de Maeztu Units of Excellence Programme (MDM-2021-001195-M), by the Fractus-UPF Chair on Tech Transfer and 6G, and by ICREA. Robert W. Heath Jr. is with the Department of Electrical and Computer Engineering, North Carolina State University, 890 Oval Dr., Raleigh, NC, 27606, USA (e-mail: rwheathjr@ncsu.edu). }
}

\author{\IEEEauthorblockN{Heedong~Do},
 {\it Member,~IEEE},
\and
\IEEEauthorblockN{Namyoon~Lee},
{\it Senior Member,~IEEE},
\and
\IEEEauthorblockN{Robert~W.~Heath~Jr.},
{\it Fellow,~IEEE},
\and
\IEEEauthorblockN{Angel~Lozano},
{\it Fellow,~IEEE}
}
\maketitle

\maketitle

\begin{abstract}
% As the satellite industry strives to meet the increasing demand for global coverage and high-speed connectivity, the need arises for wideband multiantenna communication. % in the millimeter-wave and terahertz bands.
%Striving for meeting the increasing demand for high data rate, the need arises for wideband multiantenna communication. % in the millimeter-wave and terahertz bands.
%However, for traditional phased arrays, wideband beamforming gives rise to beam squint.
With increasing frequencies, bandwidths, and array apertures, the phenomenon of beam squint arises as a serious impairment to beamforming.
% True-time-delays are a potential solution, but their analog implementation is severely limited in delay range and digital implementation requires downconversion at each element.
Fully digital arrays with true time delay per antenna element
are a potential solution, but they require downconversion at each element.
% are overly costly for satellite systems.
%As an cost-effective alternative, this paper proposes hybrid arrays.
This paper shows that hybrid arrays can perform essentially as well as digital arrays once the number of radio-frequency chains exceeds a certain threshold that is far below the number of elements. 
The result is robust, holding also for suboptimum but highly appealing beamspace architectures.
%Furthermore, this threshold remains unchanged even when implementing a beam-space architecture, which is particularly appealing. %Our comprehensive numerical study establishes the validity of this analysis in various scenarios.
% Owing to a spectrum shortage, millimeter-wave frequencies are now playing an important role in terrestrial networks since 5G roll-out, and satellite networks follow suit. It is expected that terahertz frequencies will eventually come into play. Indeed, higher frequency and satellite communications are a match made in heaven.
% Lingering coverage problem due to its poor propagation characteristic is not an issue for satellites. Also, highly directional communication could greatly alleviate the interference problem. 
% Unfortunately, such a wideband operation gives rise to beam squint resulting in reduced beamforming gain. Fully-digital arrays could be an antidote, yet expensive.
% As a cost-effective alternative, this paper concerns the use of hybrid arrays for beam squint compensation. It is shown that hybrid arrays perform almost identically to digital arrays when the number of RF chains exceeds a certain number, which is far less than the number of antenna elements. This threshold remains unchanged when employing beam-space architecture, which does not necessitate amplitude control. A comprehensive numerical study demonstrates the validity of analysis in a variety of settings.
\end{abstract}



% \fi
\IEEEpeerreviewmaketitle

\section{Introduction}

In the quest for fresh spectrum, there is much interest in millimeter wave (mmWave) and sub-terahertz bands \cite{do2021terahertz}.
Enormous amounts of bandwidth can be put into service at these frequencies, contingent on high-gain antennas to overcome the rising noise floor
%These bands offer an enormous amount of bandwidth, contingent on the deployment of high-gain antennas
\cite{boccardi2014five, andrews2014will, rangan2014millimeter, swindlehurst2014millimeter}.
% including wide bandwidth for massive throughput and narrow beam for reduced interference, both leading to major improvements in bit rate. 
%To fully realize these benefits, high-gain antennas are required at these higher frequencies.
% These benefits implicitly presume the use of high-gain antennas which can be built in compact size at these higher frequencies.
For the sake of reconfigurability, antenna arrays composed of multiple low-gain elements are preferred over a single high-gain antenna. 
% However, fully-digital antenna arrays, which require one RF chain per antenna, are currently power-hungry and require extensive baseband processing, given hundreds of antenna elements necessary to meet the link budget \cite{giordani2020satellite}.
Given that hundreds of such elements become necessary, %to close the link
fully digital arrays with one radio-frequency (RF) chain per element
are unaffordable in terms of power consumption and cost
 %are overly costly in terms of power consumption and baseband processing
 \cite{zhang2005variable, sohrabi2016hybrid, el2014spatially}.
%It is thus necessary to resort to
%As a result, %when it comes to cost,
Fully analog designs with one RF chain per array must be resorted to, with the drawback that, while the channel is frequency-dependent, analog phase shifters are frequency-independent. This results in diminished antenna gains at frequencies away from the central one, at which the array is optimized, in a phenomenon termed \emph{beam squint} \cite[Ch. 1.2]{mailloux2005phased}. % or \emph{spatial wideband effect}.

While this phenomenon can in principle be corrected with true-time-delay (TTD) beamforming \cite{mailloux1982phased, rotman2016true, longbrake2012true, cao2015advanced, mojahedian2022spatial}, the required delays for array sizes of interest are not 
 %delay range is limited. For example, a 30 cm array requires a maximum delay of only 1 ns, which is rarely
 available at mmWave frequencies when implemented at RF (see \cite[Table I]{steinweg2022hybrid} and \cite[Table I]{lin2022wideband}) and, in any event, a large chip area would be consumed. Digital true-time-delay beamforming does offer a much broader range of delays \cite[Table II]{lin2022wideband}, but again it necessitates of downconversion to baseband at each antenna element.
% Also, these approaches become obsolete when using multibeam satellites that require multiple RF chains.
% As communication technologies for terrestrial networks are developed and satellite industry follows suit soon, spatial multiplexing at the satellites would become de facto approach \cite{zheng2012generic, vazquez2016precoding, you2020massive}. These multibeam satellites necessitates the use of multiple RF chains.  
Hybrid arrays lie between the two extremes, fully analog and fully digital, and offer potentially the best of both worlds in terms of performance and cost.
The analysis of beam squint in hybrid arrays is therefore of importance, yet, in contrast with analog arrays, for which the beam-squint loss has been quantified with theoretical guarantees \cite{mailloux2005phased}, only algorithmic approaches are available for hybrid arrays \cite{liu2018space, gao2021wideband, ma2021closed, nguyen2022beam, elbir2023unified}. %which lacks theoretical guarantee and raises serious concern in computational complexity.
%Relevant contributions in the literature include \cite{cai2016effect, cai2017beamforming,wang2019beam,dovelos2021channel, dai2022delay}
%address different aspects, such as codebook design \cite{cai2016effect, cai2017beamforming} and channel estimation \cite{wang2019beam}, or explore the same problem but with an additional use of true time delay \cite{dovelos2021channel, dai2022delay}.
%\angel{Not clear what we mean here, and it might be my fault that it's unclear. Can we put in better context what these two last papers do?}
%\angel{I GUESS WE SHOULD EXPLICITLY INDICATE THAT WE WILL APPLY TTD ON THE DIGITAL PART OF OUR BEAMFORMER.} \heedong{I'm not sure that simple TTD on the digital part would work. I feel like, in general, we cannot construct MRT beamformer of propagation channel plus analog beamformer with TTD. I'll rather leave the joint use of phase shifter and TTD as a future work...}
% To address the beam squint issue in system design, fully-digital arrays are a suitable option as they are capable of frequency-dependent processing. However, when using a hybrid array with multiple RF chains per array, there may be a compromise in performance due to nonideal compensation.
In the wake of recent contributions \cite{cai2016effect, cai2017beamforming,wang2019beam,dovelos2021channel, dai2022delay}, this paper tackles the following research questions:
% aims at answering the following questions quantitatively:
% \begin{enumerate}
%     \item What is the loss in capacity when using hybrid arrays?
%     \item How many RF chains at least are required to eliminate this loss?
% \end{enumerate}
\begin{enumerate}
    \item What is the beam-squint loss in beamforming gain when using hybrid arrays in lieu of fully digital ones?
    \item Is there a number of RF chains beyond which this loss vanishes and, if so, what is that number?
%    How many RF chains are required to get rid of such loss?
\end{enumerate}

\begin{figure}
    \centering
    \includegraphics[width=0.55\linewidth]{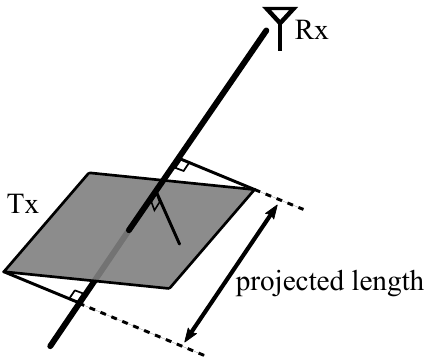}
    \caption{Projection of a planar array onto the direction of beamforming, yielding a line segment.}
    \label{fig:projectedAperture}
    \vspace*{-4mm}
\end{figure}

Remarkably, the latter question is answered in the affirmative. The minimum number of RF chains required to do away with the beam-squint loss
%In this paper, we show that a certain threshold on the number of RF chains exists where there is a diminishing return beyond this point.
emerges in a surprisingly compact closed form. For a linear array, this minimum number is simply
\be
%\frac{W}{\fc} \frac{L'}{\lambda}
\frac{\text{bandwidth}}{\text{carrier frequency}} \cdot \frac{\text{projected length}}{\text{wavelength}}
\ee
where ``projected length'' refers to the array's projection
onto the direction of beamforming. The scope extends readily to planar arrays, only with that projected length generalized to being the length of the longest segment that the array projects onto the direction of beamforming (see Fig. \ref{fig:projectedAperture}).

As the product of carrier frequency and wavelength equals the speed of light, $c$, the above can equivalently be expressed via the raw bandwidth and raw projected length, namely
\be
\frac{1}{c} \cdot \text{bandwidth} \cdot \text{projected length}.
\label{Joao}
\ee
In rather broad generality, this product of the bandwidth and the projected array length dictates the severity of the beam squint, and it directly gives the minimum number of RF chains required for squint-free beamforming with a hybrid array.
Growing bandwidths, expanding array apertures, and beamforming directions deviating from broadside, all exacerbate the squint and call for a larger number of RF chains in (\ref{Joao}), yet a hybrid array equipped with that many chains can always beamform without squint.
This behavior is asymptotic, in the sense that the squint vanishes as the product grows large,
yet it very faithfully describes the behavior for the values of interest.

For an array intending to beamform in any direction, the projected length in (\ref{Joao}) must be set to its highest value, which is 
the actual length (for a linear array) or the diagonal length (for a planar array).

The findings summarized above and expounded in the sequel are
robust, holding not only for optimum beamforming, but further for suboptimal beamspace architectures consisting of a bank of beams with a regular disposition \cite{brady2015wideband}.
The existence of a number of RF chains beyond which the squint vanishes holds even if the connectivity within the hybrid array is restricted, in the so-called hybridly-connected structure where the array is partitioned into subarrays and each subarray is connected only to a subset of RF chains. An interesting tradeoff then arises in that the number of analog phase shifters shrinks at the expense of an increased number of RF chains. Reducing both the number of analog phase shifters and RF chains leads to some residual squint being fundamentally inevitable, an aspect that we quantify for the case of a single RF chain per subarray.

The paper is organized as follows. Sec. \ref{sec:systemModel} introduces the system model and Sec. \ref{sec:beamSquintEffect} describes the phenomenon of beam squint. A simple criterion for squint-free operation is set forth in Sec. \ref{sec:approachingSquintFree}. Then, the minimum number of RF chains required for squint-free operation is quantified for linear and planar arrays in Secs. \ref{sec:linearArrays} and \ref{sec:planarArrays}, respectively. The foregoing analyses are numerically validated in Sec. \ref{sec:numericalResults}. In Sec. \ref{sec:extensions}, the results are extended to simpler architectures with restricted connectivity, and also to a multiantenna receiver. Finally, the paper concludes in Sec. \ref{sec:conclusion}.

\section{System Model}
\label{sec:systemModel}
%This section is devoted to presenting a system model that accurately describes wideband satellite systems with uniform planar array (UPA).
% An important attribute of the model analyzed in the sequel is that it features uniform planar arrays (UPAs). This does not exclude uniform linear arrays (ULAs) as ULAs can be considered a special case of UPAs.
% ; for satellite communication these are much more representative than the more analytically friendly linear arrays.

\subsection{Array Model}
\label{sec:arrayModel}
% Consider a UPA-equipped satellite communicating with a single-antenna ground station.
Consider a UPA-equipped transmitter and a single-antenna receiver.
The UPA has dimensionality $N_x\times N_y$ and aperture $L_x\times L_y$, whereby the element spacings along the respective dimensions are $d_x = \frac{L_x}{N_x}$ and $d_y=\frac{L_y}{N_y}$.
%For brevity, we adopt the zero-based numbering convention for indices.

\begin{figure}
    \centering
    \includegraphics[width=0.5\linewidth]{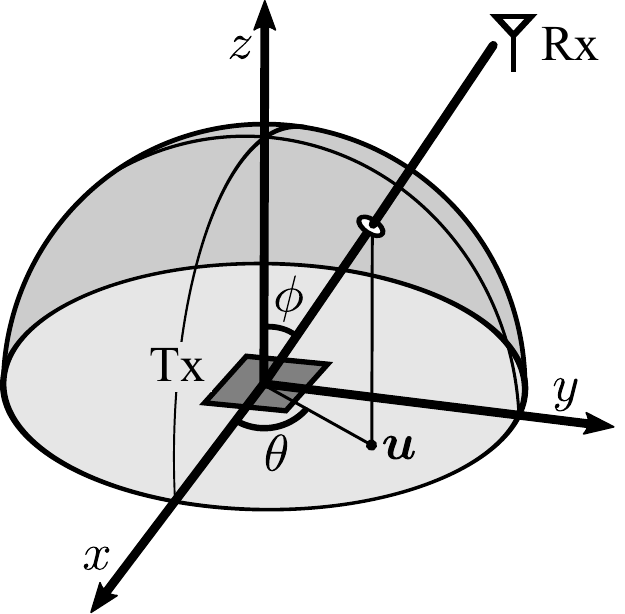}
    \caption{Considered geometry, including $\bu$ and a unit hemisphere as reference.}
    \label{fig:arrayModel}
    \vspace*{-4mm}
\end{figure}

The coordinate system is such that the $n$th transmit element ($n\in\{0,\ldots,N-1\}$ with $N=N_xN_y$) is at
\begin{align}
    \begin{bmatrix}x_n & y_n & 0\end{bmatrix}^\top  = \begin{bmatrix}d_x\big(n_x-\frac{N_x-1}{2}\big) & d_y\big(n_y-\frac{N_y-1}{2}\big) & 0\end{bmatrix}^\top
\nonumber
\end{align}
where $n_x\in\{0,\ldots,N_x-1\}$ and $n_y\in\{0,\ldots,N_y-1\}$ are the quotient and remainder of $n/N_y$.
The receiver is located at $D \begin{bmatrix}
    \sin\phi\cos\theta & \sin\phi\sin\theta & \cos\phi
\end{bmatrix}^\top$ 
where $D$ is the communication range, $\phi$ is the zenith angle, and $\theta$ is the azimuth angle (see Fig. \ref{fig:arrayModel}).
For convenience, let us henceforth define
\begin{align}
    \br_n =  \begin{bmatrix}
    x_n & y_n 
    \end{bmatrix}^\top \qquad\quad 
    \bu&=
    \begin{bmatrix}
    u_x & u_y 
    \end{bmatrix}^\top
\label{catala}
\end{align}
where $u_x = \sin\phi\cos\theta$ and $u_y = \sin\phi\sin\theta$. Note that $\bu$, also depicted in Fig. \ref{fig:arrayModel}, corresponds to the $uv$-coordinates,
handy for many applications including array processing \cite{mailloux2005phased}. It is a convenient two-dimensional projection of the unit-vector pointing to the receiver; the higher its magnitude, the further from broadside, with $\| \bu \|=0$ and $\| \bu \|=1$ respectively denoting the exact broadside and endfire directions.

The consideration of UPAs does not exclude uniform linear arrays (ULAs), as the latter are a special case of the former.
Our convention for linear arrays is to set $N_y=1$.

\subsection{Channel Model}

% We posit a pure line-of-sight connection between the satellite and the ground station and,
% for the sake of specificity, regard the satellite as the transmitter \cite[Sec. 5.3.1]{heath2018foundations}.
We posit a line-of-sight (LOS) connection.
Denoting the carrier frequency by $\fc$, and the frequency relative to it by $f\in[-\frac{W}{2},\frac{W}{2}]$ with $W$ the bandwidth, the $n$th entry of the normalized channel vector $\ba^*$ at $\fc+f$ is
\begin{align}
    [\ba^*(f)]_n = \exp \! \bigg(j2\pi \frac{\fc+f }{c}\bu^\top \br_n\bigg). \label{losChannel}
\end{align}
Such channel vector can be expressed as \cite[Eq. 1.59]{mailloux2005phased}
\begin{align}
    \ba^*(f) = \ba_{x}^*(f)\kron \ba_{y}^*(f) \label{upaChannelSeparable}
\end{align}
where $\ba_x(f)\in\bbC^{N_x}$ and $\ba_y(f)\in\bbC^{N_y}$ are vectors with entries
\begin{align}
    &[\ba_x^*(f)]_n = \exp \! \bigg(j2\pi \frac{\fc+f }{c}u_x x_n \bigg)\\
    &[\ba_y^*(f)]_n = \exp \! \bigg(j2\pi \frac{\fc+f }{c}u_y y_n \bigg).
\end{align}
% \begin{align}
%     &\ba_x^*(f) = e^{j2\pi  \frac{(\fc+f)d_x u_x}{c} \frac{N_x-1}{2}}\nonumber\\
%     &\quad \cdot\begin{bmatrix} e^{-j2\pi  \frac{(\fc+f)d_x u_x}{c}0 } & \ldots & e^{-j2\pi  \frac{(\fc+f)d_x u_x}{c}(N_x-1)} \end{bmatrix}\\
%     &\ba_{y}^*(f) = e^{j2\pi  \frac{(\fc+f)d_y u_y}{c} \frac{N_y-1}{2}}\nonumber\\
%     &\quad \cdot\begin{bmatrix} e^{-j2\pi  \frac{(\fc+f)d_y u_y}{c}0} & \ldots & e^{-j2\pi  \frac{(\fc+f)d_y u_y}{c}(N_y-1)} \end{bmatrix}
% \end{align}
The amplitude factor omitted by the normalization can be directly incorporated into the signal-to-noise ratio, giving\footnote{Large-scale effects such as atmospheric attenuation \cite{song2011present}, which is not negligible for long-range transmissions, can be easily incorporated.}
\begin{align}
    \SNR = \frac{\lambda^2 \Gt \Gr \Pt}{(4\pi D)^2 W N_0} \label{snr}
\end{align}
where $\Gt$ and $\Gr$ are the transmit and receive element gains, $\Pt$ is the total radiated power, and $N_0$ is the noise spectral density.

\subsection{Signal Model}
\label{signalModel}

A hybrid array is considered, with the number of RF chains being $\Nrf$. Denoting the analog and digital beamforming stages by
$\bW_{\sf a}\in \bbC^{N\times\Nrf}$ and $\bw_{\sf d}(f) \in \bbC^{\Nrf}$, respectively, the relation between the transmit signal $s(f)$ and the receive signal $y(f)$ is given by
\begin{align}
    y(f) = \ba^*(f)\bW_{\sf a}\bw_{\sf d}(f) s(f) + v(f) \label{initialSignalModel}
\end{align}
where $v(f)$ is a white Gaussian noise process with unit power spectral density. 
A single signal stream is transmitted, as the channel is of rank one.
%and there is no point of transceiving multiple streams from information-theoretic standpoint.

Letting $s(f)$ have unit power, the power constraint becomes
\begin{align}
    \frac{1}{W}\int_{-\frac{W}{2}}^{\frac{W}{2}} \|\bW_{\sf a}\bw_{\sf d}(f)\|^2 df = \SNR. \label{powerConstraint}
\end{align}
Introducing 
\begin{align}
    p(f) &= \|\bW_{\sf a}\bw_{\sf d}(f)\|^2 \\
    g(f) &= \frac{|\ba^*(f)\bW_{\sf a}\bw_{\sf d}(f)|^2}{\|\bW_{\sf a}\bw_{\sf d}(f)\|^2}, \label{gainDefinition}
\end{align}
we can succinctly rewrite the signal model as
\begin{align}
    y(f) = \sqrt{g(f)p(f)} s(f) + v(f) \label{finalSignalModel}
\end{align}
with the streamlined power constraint
\begin{align}
    \frac{1}{W}\int_{-\frac{W}{2}}^{\frac{W}{2}} p(f) df = \SNR.
\end{align}
Respectively, $g(f)$ and $p(f)$ can be interpreted as the beamforming gain and transmit power at frequency $f$.

% To simplify this constraint,
For a given analog beamformer, we can optimize the digital beamformer. 
To do so, we introduce the auxiliary quantity \cite[Lemma 5]{alkhateeb2016frequency}
\begin{align}
    \tilde{\bw}_{\sf d}(f) \equiv (\bW_{\sf a}^*\bW_{\sf a})^{\frac{1}{2}} \bw_{\sf d}(f), \label{auxiliary}
\end{align}
which is a natural choice from
% \begin{align}
%     \|\bW_{\sf a}\bw_{\sf d}(f)\|^2 &= \bw_{\sf d}^*(f)(\bW^*_{\sf a}\bW_{\sf a})\bw_{\sf d}(f)\\
%     &= \big\|(\bW_{\sf a}^*\bW_{\sf a})^{\frac{1}{2}}\bw_{\sf d}(f) \big\|^2.
% \end{align}
\begin{align}
    p(f) &= \bw_{\sf d}^*(f)(\bW^*_{\sf a}\bW_{\sf a})\bw_{\sf d}(f)\\
    &= \|\tilde{\bw}_{\sf d}(f) \|^2.
\end{align}
Plugging \eqref{auxiliary} into \eqref{gainDefinition}, we obtain
\begin{align}
    g(f) &= \frac{\|\ba^*(f)\bW_{\sf a}(\bW_{\sf a}^*\bW_{\sf a})^{-\frac{1}{2}}\tilde{\bw}_{\sf d}(f)\|^2}{\|\tilde{\bw}_{\sf d}(f) \|^2}\\
    &\leq \|(\bW_{\sf a}^*\bW_{\sf a})^{-\frac{1}{2}}\bW_{\sf a}^*\ba(f)\|^2 \label{gainSimplified}
\end{align}
where the upper bound follows from the Cauchy-Schwarz inequality. For any $p(f)$, such bound can attained by
\begin{align}
    \tilde{\bw}_{\sf d}(f) = \sqrt{p(f)}\frac{(\bW_{\sf a}^*\bW_{\sf a})^{-\frac{1}{2}}\bW_{\sf a}^*\ba(f)}{\|(\bW_{\sf a}^*\bW_{\sf a})^{-\frac{1}{2}}\bW_{\sf a}^*\ba(f)\|}.
\end{align}

\section{Beam Squint}
\label{sec:beamSquintEffect}

% This section concisely expounds what beam squint effect is.
% Some existing results are refined and generalized.

% \section{Analog and Hybrid Arrays With Partially-Connected Analog Network}

%\begin{figure}
%    \centering
%    \includegraphics[width=0.99\linewidth]{fejer.eps}
%    \caption{Fej\'{e}r kernel from $-\pi$ to $\pi$, which is sufficient thanks to the periodicity. \heedong{Not sure this figure is necessary.}}
%    \label{fejerGraph}
%\end{figure}

The beam squint effect is evidenced by observing the beamforming gain behavior on the space-frequency plane.
To make the dependence on both space and frequency explicit, in this section the notation $\ba(\bu,f)$ and $g(\bu,f)$ is used in lieu of $\ba(f)$ and $g(f)$ when appropriate.

With a unit-norm analog combiner $\bw_{\sf a}$ and no digital combiner,  \eqref{gainDefinition} reduces to
\begin{align}
    g(\bu,f) = |\bw_{\sf a}^*\ba(\bu,f)|^2,
\end{align}
which is often termed \emph{beam pattern} in terms of its dependence on $\bu$ \cite{mailloux2005phased}.
%where $\bw_{\sf a}$ is a normalized analog combiner satisfying $\|\bw_{\sf a}\|^2=1$.
From \eqref{losChannel},
\begin{align}
    \ba^*(\bu,f) = \ba^*\big((1+f/\fc)\bu,0\big),
\end{align}
which gives
\begin{align}
    g(\bu,f) = g\big((1+f/\fc)\bu,0\big). \label{productIsOfImportance}
\end{align}
This implies that the beam pattern at $\fc+f$ is a scaled version of that at $\fc$. Accordingly, the peak of the formed beam changes its direction with the frequency,
%(except if it is at $\bu=\begin{bmatrix}0 & 0\end{bmatrix}^\top$),
which is why the phenomenon is termed beam squint; only perfectly broadside beams are immune.
% \heedong{Good point. Perhaps, adding a figure (UPA + MRT) would perfectly work.}
% \angel{It's not crucial, but it might be a good idea to invest the time in putting together such a picture. Will be useful for your talks, posters, etc.}
This argument is valid for any array geometry and beamformer, therefore it is essentially a generalization of \cite[Lemma 2]{dai2022delay}.

As a guideline for when the squint becomes significant, the notion of 3-dB loss in beamforming gain is often used
%To quantify the intensity of the squint, the notion of 3-dB bandwidth is often used 
\cite[Ch. 1.2]{mailloux2005phased}.
%It is the smallest bandwidth $W_{\sf 3dB}$ where the beamforming gain $g(f)$ at the higher (or lower) end of the bandwidth is half of its maximum.
For ease of exposition, consider an $N$-element ULA on the $x$-axis and maximum ratio transmission (MRT), whereby $\bw_{\sf a}=\frac{\ba(0)}{\sqrt{N}}$.
%which is also called uniform illumination in phased array theory, at the center frequency.
Then,
\begin{align}
    g(f)&= \frac{|\ba^*(f)\ba(0)|^2}{N}\\
    &=\frac{1}{N}\Bigg|\sum_{n=0}^{N-1} e^{-j2\pi \frac{f d_x u_x}{c}n}\Bigg|^2\\
    &=F_{N}\bigg(\frac{2\pi d_x u_x}{c}f\bigg) \label{arrayGainMrt}
\end{align}
with $F_{N}(x) = \frac{1}{N}\Big(\frac{\sin \frac{N x}{2}}{\sin \frac{x}{2}}\Big)^{\!2}$. % is the Fej\'{e}r kernel (see Fig. \ref{fejerGraph}).
The 3-dB loss occurs when
%The 3-dB bandwidth is the smallest $W>0$ satisfying
\begin{align}
    F_{N}\bigg(\frac{\pi  d_x u_x}{c} W\bigg) = \frac{N}{2}. \label{3dBbandwidth}
\end{align}
%\heedong{How about introducing $\alpha$ here?}
Given the array length $L_x = N d_x$, the argument of $F_N(\cdot)$ above suggests defining
\begin{align}
    \alpha \equiv \frac{W}{f_{\rm c}} \frac{L_x u_x}{\lambda} 
    \label{EOS}
\end{align}
and, indeed, this product of the normalized bandwidth and normalized projected aperture---termed \emph{channel dispersion factor} in \cite{brady2015wideband}---turns out to be an excellent measure of the squint intensity with analog beamforming and plays a central role in the sequel. With it, \eqref{3dBbandwidth} is compacted into
\begin{align}
    F_{N}\bigg(\frac{\pi \alpha}{N} \bigg) = \frac{N}{2} \label{alpha3db}
\end{align}
and, as $N\rightarrow \infty$, it boils down to 
\begin{align}
    \bigg(\frac{\sin \frac{\pi\alpha}{2}}{\frac{\pi\alpha}{2}}\bigg)^{\!2} = \frac{1}{2},
\end{align}
whose numerical solution gives
\begin{align}
    \alpha_{\sf 3dB} \approx 0.886. \label{ruleOfThumb}
\end{align}
As shown in App.~\ref{3dBbandwidthProof}, $\alpha_{\sf 3dB}$
decreases with $N$ and the convergence to $0.886$ is quick, hence
this value holds virtually for any $N$; the error is only $5\%$ for $N=3$ \cite[Fig. 2.4]{hansen2009phased}.
%An analogous result for any threshold can be readily obtained.

Just like a 3-dB loss in gain maps to $\alpha_{\sf 3dB}$, any other loss value has its corresponding $\alpha$. For any such $\alpha$, as dictated by (\ref{EOS}), the larger the bandwidth, the smaller the array must be;
and the further the beam points from the broadside direction, the smaller that the bandwidth and/or the array need to be.
Since, for an exact broadside orientation, there is no beam squint, it is henceforth assumed that $\alpha >0$.

\section{Approaching Squint-Free Performance}
\label{sec:approachingSquintFree}

In this section, a simple criterion is introduced to determine whether the beamformer is approaching
%in a sense that shall become clear,
squint-free performance.

\subsection{Average Beamforming Gain}

Let us define the average beamforming gain
\begin{align}
    \gavg = \frac{1}{W}\int_{-\frac{W}{2}}^{\frac{W}{2}} g(f) df
    \label{gavgdef}
\end{align}
and, by means of the positive definite matrix
\begin{align}
    \bB = \frac{1}{W}\int_{-\frac{W}{2}}^{\frac{W}{2}} \ba(f) \ba^*(f)  df ,
     \label{matrixB}
\end{align}
further express such average gain as \cite[Eq. 20]{park2017dynamic}
\begin{align}
\label{lapro}
    \gavg&=\frac{1}{W}\int_{-\frac{W}{2}}^{\frac{W}{2}} \|\ba^*(f)\bW_{\sf a}(\bW_{\sf a}^*\bW_{\sf a})^{-\frac{1}{2}} \|^2 df\\
    % &=\frac{1}{W}\int_{-\frac{W}{2}}^{\frac{W}{2}} \tr\big( \ba^*(f)\bW_{\sf a}(\bW_{\sf a}^*\bW_{\sf a})^{-1}\bW_{\sf a}^* \ba(f)  \big)df\\
    % &=\frac{1}{W}\int_{-\frac{W}{2}}^{\frac{W}{2}} \tr\big( \bW_{\sf a}(\bW_{\sf a}^*\bW_{\sf a})^{-1}\bW_{\sf a}^* \ba(f) \ba^*(f)  \big)df\\
    &=\tr\big( \bW_{\sf a}(\bW_{\sf a}^*\bW_{\sf a})^{-1}\bW_{\sf a}^* \bB\big). \label{arrayGainPark}
\end{align}
Note that
\begin{align}
    [\bB]_{n',n} = \frac{1}{W}\int_{-\frac{W}{2}}^{\frac{W}{2}} \exp \! \bigg(j2\pi\frac{\fc+f}{c} \bu^\top(\br_n-\br_{n'})\bigg) df 
\end{align}
and, as far as the eigenvalues of $\bB$ are concerned, the term
\begin{align}
    \exp \! \bigg(j2\pi\frac{\fc}{c} \bu^\top(\br_n-\br_{n'})\bigg)
\end{align}
is irrelevant. %---it is tantamount to two unitary diagonal matrices.
Therefore, without loss of generality in terms of the eigenvalues, we can let $\fc = 0$ to obtain
\begin{align}
    [\bB]_{n',n} &= \frac{1}{W}\int_{-\frac{W}{2}}^{\frac{W}{2}} \exp \! \bigg(j2\pi \frac{f}{c} \bu^\top(\br_n-\br_{n'}) \bigg) df\\
    &=\frac{\sin \! \big( \pi W\bu^\top(\br_n-\br_{n'})/c\big)}{\pi W\bu^\top(\br_n-\br_{n'})/c}. \label{sinc}
\end{align}

\subsection{Simple Criterion}
\label{simpleCriterion}

% With fully digital beamforming, the squint can be eradicated to attain $g(f)=N$, which is equivalent to $\gavg = N$.
With fully digital beamforming, the squint can be eradicated to attain $g(f)=N$.
%The beamforming gain over $[-\frac{W}{2},\frac{W}{2}]$ can be compactly summarized with a single quantity $\gavg$
Since $g(f) \leq N$, it follows that $\gavg = N$ is a necessary and sufficient condition for $g(f)=N$.
Thus, $\gavg \approx N$ is a simple criterion to determine whether the beamformer is approaching squint-free performance.
%\angel{Might wanna emphasize that it's necessary and sufficient}
%\heedong{It can be argued using lower and upper bounds but it would make things unnecessarily complicated. Do you have better idea?}
%\angel{Perhaps I'm deceiving myself, but, since $g(f) \leq N$, if we have $\gavg = N$ then for sure $g(f)=N$. And if $g(f)=N$, then $\gavg = N$. This seems to me necessary and sufficient already.} \heedong{That's right. I was thinking of some $\epsilon$-business as approximation symbol is used!}

\subsection{Analog Beamformer Maximizing the Average Beamforming Gain}
\label{optimalBeamformer}

Given $\lambda_\ell(\bB)$ as the $\ell$th largest eigenvalue of $\bB$,
and $\bu_\ell$ as the corresponding unit-norm eigenvector, 
$\gavg$ is maximized by \cite[Prop. 1]{park2017dynamic} 
\begin{align}
    \bW_{\sf a} = 
    \begin{bmatrix}
        \bu_0 & \cdots & \bu_{\Nrf-1} 
    \end{bmatrix}
\end{align}
at
\begin{align}
    \gavg = \sum_{\ell<\Nrf} \!\! \lambda_\ell(\bB). \label{gainEigenvalue}
\end{align}
Note that 
\begin{align}
    \sum_{n}\lambda_\ell(\bB) = \tr(\bB) = N \label{eigenvalueSum}
\end{align}
consistent with the fact that an all-digital implementation ($\Nrf = N$) incurs no beam squint.

\section{Linear Arrays}
\label{sec:linearArrays}

Before embarking on the analysis with UPAs, let us first entertain the ULA case, i.e., for $N_y=1$.
%Then, in Sec. \ref{sec:planarArrays}, the UPA results will be built upon the results presented herein.

\subsection{Continuous-Aperture Representation}
From \eqref{gainEigenvalue}, the eigenvalues of $\bB$ are of great importance. For the sake of analysis, we move into the continuous realm by replacing the discrete array with a continuous one. 

For linear arrays, \eqref{sinc} reduces to
\begin{align}
    [\bB]_{n',n} &= \frac{\sin \! \big( \pi Wu_xd_x(n-n')/c\big)}{\pi Wu_xd_x(n-n')/c}\\
    &=\frac{\sin \pi\alpha \frac{n'-n}{N}}{\pi\alpha \frac{n'-n}{N}}.
\end{align}
The symmetry enables us to assume $\alpha > 0$ without loss of generality.
The continuous counterpart to $\bB$ is the integral operator \cite[Sec. V]{do2022intelligent}
\begin{align}
    \cB_\alpha: & L^2(\bbR) \rightarrow L^2(\bbR)\\
    & \,\,\,\, s(r) \,\,\,\, \mapsto \int B_\alpha(r',r) s(r) dr \nonumber
\end{align}
where 
% \begin{align}
%     B(r',r) &= [r \in A][r' \in A] \, \frac{\sin \! \big( \pi Wu_x(r-r')/c\big)}{\pi (r-r')} \label{kernelUla}
% \end{align}
% is the kernel and $A = \frac{L_x}{2}[-1,1]$ is the continuous aperture.
\begin{align}
    B_\alpha(r',r) &= [r\in I][r'\in I] \, \frac{\sin \! \big( \pi \alpha(r-r')\big)}{\pi (r-r')} \label{kernelUla}
\end{align}
is the kernel with $I = [-\frac{1}{2},\frac{1}{2}]$. Here,
% $\alpha \equiv \frac{W}{c}L_x u_x$, which plays a central role in the sequel, while
$[\cdot]$ denotes the Iverson bracket \cite{knuth1992two}
\begin{align}
    [\text{condition}] \equiv \begin{cases}
    1 &\text{the condition is true}\\
    0 &\text{otherwise}
    \end{cases}.
\end{align}
The matrix $\bB$ can be recovered from \eqref{kernelUla} by sampling at the normalized element coordinates, namely
\begin{align}
    [\bB]_{n',n} = \frac{1}{\alpha}B_\alpha\bigg(\frac{r_{n'}}{L_x}, \frac{r_n}{L_x}\bigg) ,
\end{align}
where the normalization by $\frac{1}{\alpha}$ is convenient, as evidenced in subsequent sections.
%in order to make the analysis (to be introduced in the subsequent sections) simpler. 
In terms of eigenvalues, the continuous representation becomes exact as the spacings vanish, namely \cite[Sec. V-C]{do2022intelligent}
\begin{align}
    \frac{\alpha}{N} \lambda_\ell(\bB) \rightarrow \lambda_\ell(\cB_\alpha)
    \label{eigenvalueRelation}
\end{align}
in two-norm as $N$ approaches infinity.

For any $\alpha$, which measures the squint intensity that would be experienced with purely analog beamforming (recall Sec. \ref{sec:beamSquintEffect}),
the squint-free condition with hybrid beamforming, $\sum_{\ell<\Nrf}\lambda_\ell(\bB) \approx N$,
translates to
\begin{align}
    \sum_{\ell<\Nrf} \!\! \lambda_\ell(\cB_\alpha) \approx  \alpha \label{squintFreeCondition}
\end{align}
and the counterpart to the all-digital extreme in \eqref{eigenvalueSum} is $\sum_{\ell=0}^{\infty}\lambda_\ell(\cB_\alpha) = \alpha$.

The continuous-aperture representation introduced herein is an analytically friendly proxy to the discrete model,
%which is different from the philosophy of holographic MIMO \cite{pizzo2020spatially}. A practical relevance of the analysis relying on continuous-aperture representation will be
and its practical relevance is confirmed by thorough numerical studies.

\subsection{Eigenvalue Behavior}

% The eigenvalues of $\cB_\alpha$ are well studied \cite{slepian1961prolate}. In particular, they are:
% \begin{enumerate}
%     % \item A function of $\alpha$; in this regard, we hereafter use the notation $\lambda_n(\cB_\alpha)$ in lieu of $\lambda_n(\cB)$.
%     \item Bounded above by 1.
%     \item Asymptotically polarized into two levels; precisely, as $\alpha\rightarrow \infty$,
%     \begin{align}
%         |\{\ell:\lambda_\ell(\cB_\alpha)>\epsilon\}| = \alpha +\cO(\log \alpha) \label{polarizationFormal}
%     \end{align}
%     for any $\epsilon\in(0,1)$.
% \end{enumerate}
The eigenvalues of $\cB_\alpha$ are well studied \cite{slepian1961prolate}. In particular, they are bounded above by 1 and asymptotically polarized into two levels. Precisely, as $\alpha\rightarrow \infty$,
\begin{align}
|\{\ell:\lambda_\ell(\cB_\alpha)>\epsilon\}| = \alpha +\cO(\log \alpha) \label{polarizationFormal}
\end{align}
for any $\epsilon\in(0,1)$.
Put another way, for sufficiently large $\alpha$, we can informally write
\begin{align}
    (\lambda_0(\cB_\alpha),\lambda_1(\cB_\alpha),\ldots) \approx (\underbrace{1,\ldots,1}_{\approx \alpha},0,\ldots). \label{polarizationInformal}
\end{align}

\subsection{Number of Required RF chains}

Let us consider the asymptotic regime where $\alpha\rightarrow \infty$ and $\Nrf = \lceil p\alpha \rceil$ with $p>0$ a constant. Then,
\begin{align}
    \sum_{\ell< \Nrf}\lambda_\ell(\cB_\alpha) = \min(p,1) \alpha + o(\alpha), \label{ulaSquintFree}
\end{align}
which (see App. \ref{ulaProof}) is a consequence of \eqref{polarizationInformal}. Therefore,
\begin{align}
    \sum_{\ell < \Nrf}\lambda_\ell(\cB_\alpha) = \alpha + o(\alpha)
\end{align}
if $p\geq 1$. Recalling the squint-free condition in \eqref{squintFreeCondition}, this implies that, asymptotically, $\alpha$ RF chains are needed to approach squint-free performance. 
% \heedong{Slightly expanded.}
% \angel{Good. But the summations in this subsection are over $n \geq \Nrf$ rather than $n < \Nrf$}

\subsection{Beamspace Architecture}
\label{beamSpaceLinear}

An instance of the above general ULA result had been empirically observed 
in the context of beamspace MIMO \cite{brady2015wideband}, namely that the beam squint can be mitigated with $\lceil\alpha\rceil$ RF chains and a simple analog network.
The $\ell$th column of $\bW_{\sf a}$ is then the MRT beamformer $\frac{\ba(f_\ell)}{\sqrt{N}}$ for the relative frequency
\begin{align}
    f_\ell = \frac{c}{u_x d_x}\cdot \frac{\ell-\frac{\Nrf-1}{2}}{N} =\frac{\ell-\frac{\Nrf-1}{2}}{\alpha} W, \label{beamSpaceFrequency}
\end{align}
% Explicitly,
% \begin{align}
%     [\ba(f_\ell)]_n
%     &= \exp \! \bigg(\!-j2\pi \frac{\fc+f_\ell }{c}u_x x_n\bigg)\\
%     &= \exp \! \bigg(\!-j2\pi \frac{\fc}{c}u_x x_n\bigg)\nonumber\\
%     &\quad\cdot \exp\!\bigg(-j2\pi \frac{(\ell-\frac{\Nrf-1}{2})(n-\frac{N-1}{2})}{N}\bigg) ,
% \end{align}
% \begin{align}
%     %&
%     \frac{1}{\sqrt{N}} \! \begin{bmatrix}
%     e^{-j2\pi\frac{\fc}{c}u_x x_0} \\ \vdots \\ e^{-j2\pi\frac{\fc}{c}u_x x_{N-1}}
%     \end{bmatrix} %\nonumber\\
%     %&\qquad\qquad\qquad
%     \!\circ\!
%     \begin{bmatrix}
%     e^{-j2\pi(\ell-\frac{\Nrf-1}{2})(0-\frac{N-1}{2})/N} \\ \vdots \\ e^{-j2\pi(\ell-\frac{\Nrf-1}{2})((N-1)-\frac{N-1}{2})/N}
%     \end{bmatrix},
%     \label{beamSpaceColumn}
% \end{align}
such that $W$ is segregated into $\Nrf$ subbands spaced by $\frac{W}{\alpha}$.
Intuitively, this narrower subbands are less susceptible to beam squint; in fact, they are dimensioned such that the squint vanishes asymptotically over each of them, with the digital stage of the beamformer taking care of the rest.

% \angel{Check this version:}
A dual interpretation, recalling \eqref{productIsOfImportance}, is that the $\ell$th analog beamformer points (at its own central frequency) in direction $(1+\frac{f_\ell}{\fc})u_x$. At the analog stage, we are therefore faced with a bank of beams arranged regularly in $u_x$, hence the
beamspace denomination.

%\heedong{Recalling \eqref{productIsOfImportance}, this beamformer points the direction $(1+\frac{f_\ell}{\fc})\bu$ at the center frequency.} \angel{You mean at the center frequency of the $\ell$th beam?} \heedong{Right.} \heedong{I think it essentially explains why this architecture is termed ``beamspace''. Without this, readers may wonder...}

%This points to $\Nrf \geq \alpha$, which is verified shortly.

% Mathematically, beam-space MIMO is essentially a hybrid precoding architecture where the $n$th column of $\bW_{\sf a}$ is
% \begin{align}
%     %&
%     \frac{1}{\sqrt{N}} \! \begin{bmatrix}
%     e^{-j2\pi\frac{\fc}{c}u_x x_0} \\ \vdots \\ e^{-j2\pi\frac{\fc}{c}u_x x_{N-1}}
%     \end{bmatrix} %\nonumber\\
%     %&\qquad\qquad\qquad
%     \!\circ\!
%     \begin{bmatrix}
%     e^{-j2\pi(n-\frac{\Nrf-1}{2})(0-\frac{N-1}{2})/N} \\ \vdots \\ e^{-j2\pi(n-\frac{\Nrf-1}{2})((N-1)-\frac{N-1}{2})/N}
%     \end{bmatrix},
%     \label{beamSpaceColumn}
% \end{align}
% where $\circ$ denotes the Hadamard product. It can be interpreted as the MRT beamformer for the relative frequency
% \begin{align}
%     \frac{c}{u_x d_x}\cdot \frac{n-\frac{\Nrf-1}{2}}{N} =\frac{n-\frac{\Nrf-1}{2}}{\alpha} W. \label{beamSpaceFrequency}
% \end{align}

% The first term therein is based on the channel at the center frequency while the second term corrects the beam squint.

\begin{figure}
    \centering
    \includegraphics[width=0.99\linewidth]{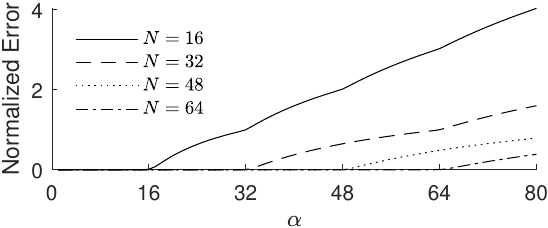}
    \caption{Validity of the continuous-aperture representation.}
    \label{fig:approximationError}
    \vspace*{-4mm}
\end{figure}

As the columns are orthonormal, \eqref{lapro} gives
\begin{align}
    \frac{\gavg}{N}
    &=\frac{1}{NW}\int_{-\frac{W}{2}}^{\frac{W}{2}} \|\ba^*(f)\bW_{\sf a}\|^2df\\
    &=\sum_{\ell<\Nrf}\frac{1}{W}\int_{-\frac{W}{2}}^{\frac{W}{2}}\frac{|\ba^*(f)\ba(f_\ell)|^2}{N^2}df. \label{beamspaceDerivation}
\end{align}
To again move into the continuous-aperture realm, we let $N \to \infty$ while retaining the aperture, whereby
\begin{align}
    \frac{|\ba^*(f)\ba(f_\ell)|^2}{N^2} & =\Bigg(\frac{\sin\big(\pi\big(\ell-\frac{\Nrf-1}{2}-\frac{f}{W}\alpha \big)\big)}{N\sin\big(\pi\big(\ell-\frac{\Nrf-1}{2}-\frac{f }{W}\alpha \big)/N\big)}\Bigg)^{\! 2} \nonumber \\
    & \rightarrow \Bigg(\frac{\sin\!\big(\pi\big(\ell-\frac{\Nrf-1}{2}-\frac{f} {W}\alpha\big)\big)}{\pi\big(\ell-\frac{\Nrf-1}{2}-\frac{f}{W}\alpha \big)}\!\Bigg)^{\!2} \label{smallBandwidthPerRfChain}
\end{align}
%\angel{Not obvious that the argument of the $\sin()$ is always small, we should say something about this...} \heedong{Perhaps explicitly writing $\ell<\Nrf$ would work? (it's not $\ell<N$)}
%which corresponds to approximating the array with a continuous aperture. That is, this approximation becomes exact as $N\rightarrow \infty$ while keeping the aperture (i.e., $d_x = \frac{L_x}{N}$) and the number of RF chains $\Nrf$.
%Plugging it into \eqref{beamspaceDerivation}, 
and
\begin{align}
    \frac{\gavg}{N} &\rightarrow \sum_{\ell<\Nrf}\frac{1}{W}\int_{-\frac{W}{2}}^{\frac{W}{2}}\Bigg(\frac{\sin\big(\pi\big(\ell-\frac{\Nrf-1}{2}-\frac{f}{W}\alpha\big)\big)}{\pi\big(\ell-\frac{\Nrf-1}{2}-\frac{f}{W}\alpha\big)}\Bigg)^{\!2} df\nonumber\\
    &= \frac{1}{\alpha}\sum_{\ell<\Nrf}\int_{\ell-\frac{\Nrf-1}{2}-\frac{\alpha}{2}}^{\ell-\frac{\Nrf-1}{2}+\frac{\alpha}{2}}\bigg(\frac{\sin \pi t}{\pi t}\bigg)^{\!2} dt \label{beamspaceDerivationContinued}
\end{align}

Let us again consider the asymptotic regime, $\alpha\rightarrow \infty$ and $\Nrf = \lceil p\alpha \rceil$ with $p>0$ a constant. It is shown in App. \ref{beamspaceProof} that %the right-hand side of
\eqref{beamspaceDerivationContinued} converges to $\min(p,1)$. The result is thus identical to its counterpart with the optimum eigenvectors (recall \eqref{ulaSquintFree}), demonstrating the asymptotic optimality of the beamspace architecture. 

This architecture is computationally efficient and compelling in that no amplitude tapering is required. On top of that, the normalized response at the $\ell$th RF chain, given in \eqref{smallBandwidthPerRfChain}, concentrates around $f_\ell$ 
such that having a much smaller bandwidth at each RF chain incurs negligible loss. With a bandwidth of $\frac{2}{\alpha}W$ per RF chain, the average beamforming gain becomes
\begin{align}
    \int_{-1}^1 \bigg(\frac{\sin \pi t}{\pi t}\bigg)^{\!2} dt \approx 0.902,
\end{align}
which falls short by only $0.45$ dB. Doubling the bandwidth per RF chain shrinks this deficit to $0.22$ dB. 

\begin{figure}
    \centering
    \includegraphics[width=0.99\linewidth]{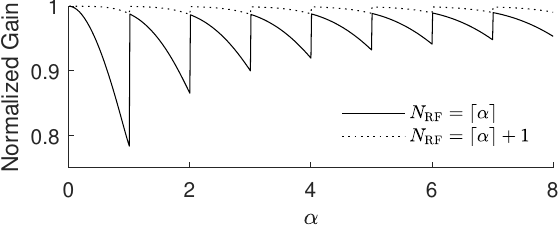}
    \caption{Normalized gain with respect to $\alpha$. The sawtooth behavior, caused by the ceiling function on the number of RF chains, quickly abates.}
    \label{fig:gainLinear}
    \vspace*{-4mm}
\end{figure}

\subsection{Remarks}

The analysis entails two limits, infinitesimal element spacing and infinite aperture-bandwidth product, and care should be exercised to preserve the relevance of the analysis to regimes of interest \cite{dohler2011phy}.

Recalling \eqref{eigenvalueRelation}, Fig. \ref{fig:approximationError} depicts the normalized approximation error for the first limit\footnote{$\lambda_\ell(\cB_\alpha)$ can be computed via the software package {\fontfamily{qcr}\selectfont chebfun2} \cite{townsend2013extension}.}
\begin{align}
    \frac{\sum_\ell  \left(\frac{\alpha}{N} \lambda_\ell(\bB) - \lambda_\ell(\cB_\alpha) \right)^2}{\sum_\ell \lambda^2_\ell(\cB_\alpha)} .
\end{align}
The figure evidences that the continuous-aperture representation is fully valid provided that $\alpha \leq N$, a condition clearly fulfilled for any scenario of interest because
\begin{align}
    \alpha = N \frac{W}{\fc} \frac{d_x u_x}{\lambda}  \ll N.
\end{align}
This leaves the second limit, base on which the analysis is valid as long as $\alpha$ is minimally large. 
%Interestingly, even for small $\alpha$ where the polarization is not holding, the result of analysis is still valid.
Pleasingly, even for very small $\alpha$ this turns out to be the case. 
Fig. \ref{fig:gainLinear} depicts the normalized gain $\big(\sum_{\ell<\Nrf}\lambda_\ell(\cB_\alpha)\big)/{\alpha}$ with respect to $\alpha$, confirming that, for virtually any $\alpha$, squint-free performance can be virtually achieved with $\lceil \alpha \rceil+1$ RF chains.

% One can think of \eqref{beamSpaceColumn} as the function
% \begin{align}
%     \exp\bigg(\!-j2\pi\bigg(\frac{\fc}{c}u_x r + \bigg(n-\frac{\Nrf-1}{2}\bigg)\frac{r}{L_x}\bigg)\bigg) \label{beamspaceUlaContinuous}
% \end{align}
% sampled at $\{r_n\}$. This is a useful interpretation when extending the architecture to UPAs.

\section{Planar Arrays}
\label{sec:planarArrays}

%\heedong{Significantly simplified thanks to the introduction of $\balpha$.}
In this section, the result is extended to planar arrays; this generalization is not straightforward in that a naive separable architecture makes an inefficient use of the RF chains. For planar arrays, the intensity of beam squint can be measured by 
\begin{align}
    \balpha &\equiv \begin{bmatrix}\alpha_x \\ \alpha_y \end{bmatrix}= \frac{W}{\fc} \begin{bmatrix}\frac{L_x u_x}{\lambda} \\ \frac{L_y u_y}{\lambda} \end{bmatrix},  \label{alphaPlanar}
\end{align}
which is a natural extension of $\alpha$ for linear arrays.
% Unlike for ULAs, though, for which the eigenvalue behavior is well-investigated in Sec. \ref{sec:linearArrays}, similar characterizations do not exist for UPAs.

% From \eqref{gainEigenvalue}, the eigenvalues of $\bB$ are of great importance. 
% Unlike for ULAs, though, for which the eigenvalue behavior is well-investigated [CITE ULA PAPER], similar characterizations do not exist for UPAs.
% To approximate such characterization, we move into the continuous realm by replacing the discrete array with a continuous one.
%To this end, we approximate the eigenvalues of $\bB$ by replacing the discrete antenna array into the continuous one. 

\subsection{Separable Architecture}
\label{separableArchitecture}

Motivated by \eqref{upaChannelSeparable}, separable beamformers are often considered where \cite[Ch. 1.2]{mailloux2005phased}
\begin{align}
    \bW_{\sf a} = \bW_{x}\kron \bW_{y}    
\end{align}
with $\bW_{x}\in\bbC^{N_x\times N_{{\rm RF},x}}$ and $\bW_{y}\in\bbC^{N_y\times N_{{\rm RF},y}}$, and with $\Nrf = N_{{\rm RF},x}N_{{\rm RF},y}$.
Then,
\begin{align}
    g(f) &= \big\|\big((\bW_{ x}^*\kron \bW_{y}^*)(\bW_{x}\kron \bW_{y})\big)^{-\frac{1}{2}} \nonumber\\
    &\qquad\qquad\qquad \cdot (\bW_{ x}^*\kron \bW_{y}^*) (\ba_x(f)\kron \ba_y(f))\big\|^2\\
    &= \big\|(\bW_{x}^*\bW_{x})^{-\frac{1}{2}} \bW_{x}^* \ba_x(f) \nonumber\\
    &\qquad\qquad\qquad \kron(\bW_{y}^*\bW_{y})^{-\frac{1}{2}} \bW_{ y}^* \ba_y(f)\big\|^2\\
    &= \big\|(\bW_{x}^*\bW_{x})^{-\frac{1}{2}} \bW_{ x}^* \ba_x(f)\big\|^2 \nonumber \\
    &\qquad\qquad\qquad \cdot \big\|(\bW_{ y}^*\bW_{y})^{-\frac{1}{2}} \bW_{y}^* \ba_y(f)\big\|^2\\
    &= g_x(f)g_y(f)
\end{align}
given
\begin{align}
g_x(f) & = \big\|(\bW_{x}^*\bW_{x})^{-\frac{1}{2}} \bW_{x}^* \ba_x(f)\big\|^2 \\
%\end{equation}
%and
%\begin{equation}
g_y(f) & = \big\|(\bW_{y}^*\bW_{ y})^{-\frac{1}{2}} \bW_{y}^* \ba_y(f)\big\|^2.
\end{align}
% \begin{align}
%     g_x(f) &= \big\|(\bW_{x}^*\bW_{x})^{-\frac{1}{2}} \bW_{x}^* \ba_x(f)\big\|^2 \nonumber\\
%     g_y(f) &= \big\|(\bW_{y}^*\bW_{ y})^{-\frac{1}{2}} \bW_{ y}^* \ba_y(f)\big\|^2.
% \end{align}
It follows that (see App. \ref{separableProof})
% \begin{align}
%     \gavg = \frac{1}{W}\int_{-\frac{W}{2}}^{\frac{W}{2}} g_x(f)g_y(f) df
%     & \leq \min(N_x g_{{\sf avg},y},N_y g_{{\sf avg},x}) \nonumber\\
%     &= N\min \! \bigg(\frac{g_{{\sf avg},x}}{N_x}, \frac{g_{{\sf avg},y}}{N_y}\bigg) \label{separableBound}
% \end{align}
\begin{align}
    \frac{\gavg}{N} \leq  \min \! \bigg(\frac{g_{{\sf avg},x}}{N_x}, \frac{g_{{\sf avg},y}}{N_y}\bigg) \label{separableBound}
\end{align}
where $g_{{\sf avg},x} \!=\! \frac{1}{W}\!\int_{-\frac{W}{2}}^{\frac{W}{2}} g_x(f) df$ and $g_{{\sf avg},y} \!=\! \frac{1}{W}\!\int_{-\frac{W}{2}}^{\frac{W}{2}} g_y(f) df$.
% \begin{align}
%     g_{{\sf avg},x} &= \frac{1}{W}\int_{-\frac{W}{2}}^{\frac{W}{2}} g_x(f) df\\
%     g_{{\sf avg},y} &= \frac{1}{W}\int_{-\frac{W}{2}}^{\frac{W}{2}} g_y(f) df.
% \end{align}
% Also, as shown in App. \ref{separableBoundProof}, we also have
% \begin{align}
%     \gavg \geq N\bigg(\frac{g_{{\sf avg},x}}{N_x} + \frac{g_{{\sf avg},y}}{N_y} -1\bigg).
% \end{align}

% For $\theta = 0$ and $\theta = \pi$,
% $\ba_y(f)$ is constant and $g_{{\sf avg},y} = N_y$, indicating no squint in the $y$-direction.
% The setting reduces to that of a uniform linear array (ULA), for which it is established in [CITE ULA PAPER] that $\frac{L_x \sin\phi}{c} W$ RF chains are required to approach squint-free performance. An analogous result applies, in the other dimension, for $\theta = \frac{\pi}{2}$ and $\theta = \frac{3\pi}{2}$. 

% More generally, 
% Whenever $u_x \neq 0$ and $u_y \neq 0$ in (\ref{catala}),
Attaining $\gavg \approx N$ with a separable structure entails $g_{{\sf avg},x}\approx N_x$ and $g_{{\sf avg},y}\approx N_y$.
For $u_x \neq 0$ and $u_y \neq 0$ in (\ref{catala}), asymptotically in the array aperture and/or the bandwidth it holds that $ N_{{\rm RF},x} \geq \alpha_x$ and $N_{{\rm RF},y} \geq \alpha_y$
% \begin{align}
%     N_{{\rm RF},x} \geq \alpha_x \qquad\quad N_{{\rm RF},y} \geq \alpha_y 
% \end{align}
as per Sec. \ref{sec:linearArrays}.
%to attain the squint-free performance.
Thus,
\begin{align}
    \Nrf \geq \alpha_x\alpha_y = \frac{L_x L_y u_x u_y}{c^2} W^2 , %\cdot \frac{L_y u_y}{c} W,
\end{align}
which is \emph{quadratic} in the diagonal dimensions of the array and also quadratic in the bandwidth.
%This natural generalization highly overestimates the required number of RF chains.
In the sequel, it is shown that a nonseparable beamformer can greatly reduce this number.

\subsection{Continuous-Aperture Representation}

Again resorting to a continuous-aperture representation, the counterpart to $\bB$ is the integral operator 
\begin{align}
    \cB_{\balpha}: & L^2(\bbR^2) \rightarrow L^2(\bbR^2)\\
    & \,\,\,\, s(\br) \,\,\,\, \mapsto \int B_{\balpha}(\br',\br) s(\br) d\br \nonumber
\end{align}
where 
% \begin{align}
%     B(\br',\br) &= [\br \in A][\br' \in A] \frac{\sin \! \big( \pi W\bu^\top(\br-\br')/c\big)}{\pi \bu^\top(\br-\br')/\sin\phi} \label{kernelUpa}
% \end{align}
% is the kernel and $A = \frac{L_x}{2}[-1,1]\times \frac{L_y}{2}[-1,1]$ is the continuous aperture. 
\begin{align}
    &B_{\balpha}(\br',\br) = \big[\br \in I^2 \big]\big[\br' \in I^2 \big] \frac{\sin \! \big( \pi \balpha^\top(\br-\br')\big)}{\pi \balpha^\top(\br-\br')/\|\balpha\|} \label{kernelUpa}
\end{align}
is the kernel.
The matrix $\bB$ can again be recovered from \eqref{kernelUpa} by sampling at the normalized element coordinates, precisely
\begin{align}
    [\bB]_{n',n} = \frac{1}{\|\balpha\|}B_{\balpha}\Bigg(\begin{bmatrix}
        \frac{x_{n'}}{L_x}\\ \frac{y_{n'}}{L_y}
    \end{bmatrix}, \begin{bmatrix}
        \frac{x_{n}}{L_x}\\ \frac{y_{n}}{L_y}
    \end{bmatrix}\Bigg).
\end{align}
In terms of eigenvalues, the continuous representation becomes exact as the spacings vanish, namely \cite[Sec. V-C]{do2022intelligent}
\begin{align}
    \frac{\|\balpha\|}{N} \lambda_\ell(\bB) \rightarrow \lambda_\ell(\cB_{\balpha})
\end{align}
in two-norm as $N_x$ and $N_y$ grow large with a fixed ratio. The squint-free condition is 
\begin{align}
    \sum_{\ell<\Nrf} \!\! \lambda_\ell(\cB_{\balpha}) \approx \|\balpha\|.
\end{align}

\subsection{A Bag of Tricks}

The operator $\cB$ obtained from the continuous-aperture representation is still ill-suited to analysis.
To enable further progress, some handy results are set forth in this subsection; they allow capitalizing on ULAs asymptotics.

Let us construct a rotation matrix $\bR\in\bbR^{2\times 2}$ whose first row is $\frac{\balpha}{\|\balpha\|}$, and then rotate the axes into
\begin{align}
    \tilde{\br} = 
    \begin{bmatrix}
        \tilde{r}_x & \tilde{r}_y
    \end{bmatrix}^\top =  \bR \br \qquad\quad
    \tilde{\br}' =
    \begin{bmatrix}
        \tilde{r}_x' & \tilde{r}_y'
    \end{bmatrix} = \bR \br' . \label{changeOfVariables}
\end{align}
The right-hand side of \eqref{kernelUpa} can then be recast as
\begin{align}
    \big[\tilde{\br} \in \bR I^2\big]
    \big[\tilde{\br}' \in \bR I^2\big] \frac{\sin \! \big( \pi \|\balpha\| (\tilde{r}_x-\tilde{r}_x') \big)}{\pi (\tilde{r}_x-\tilde{r}_x')}. \label{changeOfVariableViaRotation}
\end{align}
% \begin{figure}
%     \centering
%     \includegraphics[width=0.9\linewidth]{example-image}
%     \caption{Visualization of the bounds; checkerboard for upper bound and polka dots for lower bound.}
%     \label{trapezoidVisualization}
% \end{figure}
Defining the kernel
\begin{align}
    & \tilde{B}_{\balpha}(\tilde{r}_x',\tilde{r}_x)
    = \bigg(\int \big[ \tilde{\br} \in \bR I^2\big] d\tilde{r}_y\bigg)^{\!1/2}\nonumber\\
    &\qquad\; \cdot \bigg(\int \big[\tilde{\br}' \in \bR I^2\big] d\tilde{r}_y'\bigg)^{\! 1/2} \, \frac{\sin \! \big( \pi \|\balpha\| (\tilde{r}_x-\tilde{r}_x') \big)}{\pi (\tilde{r}_x-\tilde{r}_x')} \label{dimensionalityReduction}
\end{align}
and the corresponding operator $\tilde{\cB}_{\balpha}: L^2(\bbR) \rightarrow L^2(\bbR)$, it is shown in App. \ref{simplificationProof} that the singular values of $\tilde{\cB}_{\balpha}$ and $\cB_{\balpha}$ coincide.
If the factors
\begin{align}
    \bigg(\int \big[ \tilde{\br} \in \bR I^2\big] d\tilde{r}_y\bigg)^{\! 1/2} \qquad \bigg(\int \big[\tilde{\br}' \in \bR I^2\big] d\tilde{r}_y'\bigg)^{\! 1/2}
\end{align}
are replaced by indicator functions, the above boils down to the ULA kernel. This motivates bounding these factors with scalar multiples of the indicator function and invoking the Courant min-max theorem.
% \begin{figure*}
% \begin{align}
%     &\bigg(-\frac{L_x|\cos\theta|+L_y|\sin\theta|}{2},0\bigg), \bigg(-\frac{\big| L_x|\cos\theta|-L_y|\sin\theta|\big|}{2}, \min\bigg(\frac{L_x}{|\sin\theta|},\frac{L_y}{|\cos\theta|}\bigg) \bigg), \nonumber\\
%     &\bigg(\frac{\big| L_x|\cos\theta|-L_y|\sin\theta|\big|}{2}, \min\bigg(\frac{L_x}{|\sin\theta|},\frac{L_y}{|\cos\theta|}\bigg) \bigg), \bigg(\frac{L_x|\cos\theta|+L_y|\sin\theta|}{2},0\bigg). \label{trapezoidPoints}
% \end{align}
% \hrulefill
% \end{figure*}
Precisely, $\int \big[ \tilde{\br} \in \bR I^2\big] d\tilde{r}_y$
is a piecewise linear function connecting the points
% \begin{align}
%     &\bigg(-\frac{L_1}{2},0\bigg), \bigg(-\frac{L_2}{2}, L_3 \bigg), \bigg(\frac{L_2}{2}, L_3 \bigg), \bigg(\frac{L_1}{2},0\bigg)
% \end{align}
\begin{align}
    &\big(-L_1/2,0\big), \big(-L_2/2, L_3 \big), \big(L_2/2, L_3 \big), \big(L_1/2,0\big)
\end{align}
given
% \begin{align}
% L_1 & = \frac{|\alpha_x|+|\alpha_y|}{\|\balpha\|} \\
% L_2 & = \frac{||\alpha_x|-|\alpha_y||}{\|\balpha\|} \\
% L_3 & = \frac{\|\balpha\|}{\max(|\alpha_x|,|\alpha_y|)}
% \end{align}
$L_1 = (|\alpha_x|+|\alpha_y|)/\|\balpha\|$,
$L_2 = \big||\alpha_x|-|\alpha_y|\big|/\|\balpha\|$, and
$L_3 = \|\balpha\|/\max(|\alpha_x|,|\alpha_y|)$.
Also, it can be bounded as
\begin{align}
    &\delta L_3\big[ \tilde{r}_x\in ((1-\delta)L_1 + \delta L_2)I \big]\nonumber\\
    &\qquad\qquad\leq  \int \big[ \tilde{\br} \in \bR I^2\big] d\tilde{r}_y \leq  L_3\big[\tilde{r}_x\in L_1 I \big]  \label{Messi}
\end{align}
where $\delta\in[0,1]$ is a constant to be determined.

Using (\ref{Messi}) and the fact that $\cB$ and $\tilde{\cB}$ share the same eigenvalues, it is shown in App. \ref{minMaxProof} that
\begin{align}
    \delta L_3 \lambda_\ell(\cB_{\alpha^{\sf lo}}) \leq  \lambda_\ell(\cB_{\balpha}) \leq  L_3\lambda_\ell(\cB_{\alpha^{\sf up}}), \label{sandwich}
\end{align}
where $\alpha^{\sf lo } = \|\balpha\| \big((1-\delta)L_1+\delta L_2\big)$ and $\alpha^{\sf up} = \|\balpha\| L_1$.

\begin{figure}
    \centering
    \includegraphics[width=0.5\linewidth]{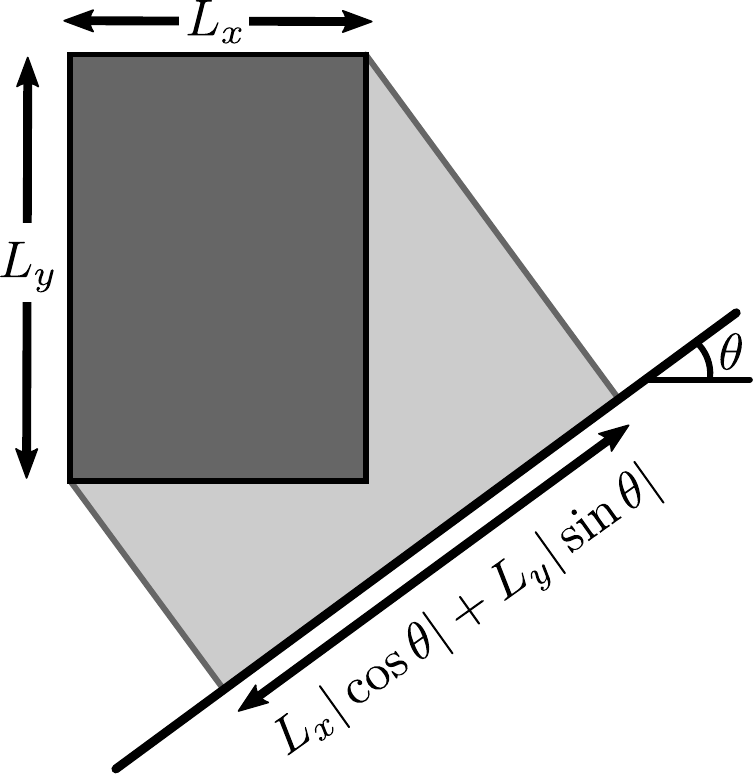}
    \caption{Projection of the UPA onto the line parallel to $\begin{bmatrix}\cos\theta & \sin\theta\end{bmatrix}^\top$.
    %\heedong{Do we need this figure? I think Fig. \ref{fig:projectedAperture} explains the concept fairly well. What we need is a slight tweak in explanation that $(L_x |\cos\theta| + L_y |\sin\theta|)\sin\phi$ is the projected length in Fig. \ref{fig:projectedAperture}.}
    }
    \label{fig:projection}
    \vspace*{-4mm}
\end{figure}

\subsection{Number of Required RF Chains}
% Let us consider the bandwidth $r_1 W$, the aperture $r_2 L_x \times r_2L_y$, and the number of RF chains
% \begin{align}
%     \Nrf = \big\lceil p \alpha^{\sf up} \big\rceil ,
% \end{align}
% where $r_1, r_2, p>0$ are constants, and (\ref{alphas}) is thereby generalized to
% $
% \alpha^{\sf up}=\frac{(r_1W)\sin\phi}{c}(r_2 L_1).
% $
% Similarly to $\lambda_n(\alpha^{\sf lo})$ and $\lambda_n(\alpha^{\sf up})$, a simple change of variables evidences that $\lambda_n(\cB)/r_2$ depend on $r\equiv r_1r_2$, not individually on $r_1$ and $r_2$. This implies that the normalized beamforming gain
% \begin{align}
%     \frac{\sum_{n < \Nrf} \lambda_n(\cB)}{\sum_{n} \lambda_n(\cB)} = \frac{\sum_{n < \Nrf} \lambda_n(\cB)/r_2}{\sum_{n} \lambda_n(\cB)/r_2}
% \end{align}
% depends only on $r$. Without loss of generality, let $r_1 = r$ and $r_2 =1$, and consider the asymptotic regime $r\rightarrow \infty$.

% Armed with \eqref{sandwich}, it is shown in App. that
% \begin{align}
%     \frac{\sum_{n \geq \Nrf} \lambda_n(\cB)}{\sum_{n} \lambda_n(\cB)} = o(1) \label{upaSquintFree}
% \end{align}
% ---this,
% %from Sec. \ref{sec:approachingSquintFree} that
% recall, ensures no beam squint---if and only if $p\geq 1$.
% It follows that
% \begin{align}
%     \alpha^{\sf up} =\frac{W\sin\phi}{c}\big( \text{array length projected on }(\cos\theta, \sin\theta)  \big)
% \end{align}
% is the minimum number of RF chains that asymptotically attain squint-free performance. It is worth noting that the analysis can be readily generalized to other array topologies.

Let us consider the bandwidth $r_1 W$ and the aperture $r_2 L_x \times r_2L_y$ where $r_1, r_2>0$ are constants. Then,
\begin{align}
    \balpha =\frac{rW}{\fc}\begin{bmatrix}\frac{L_x u_x}{\lambda} \\ \frac{L_y u_y}{\lambda} \end{bmatrix}
\end{align}
depends on $r\equiv r_1r_2$, not individually on $r_1$ and $r_2$.
With the number of RF chains at
\begin{align}
    \Nrf = \big\lceil p \alpha^{\sf up} \big\rceil,
\end{align}
let us once more consider the regime $r\rightarrow \infty$.
Armed with \eqref{sandwich}, it is shown in App. \ref{upaSquintFreeProof} that the no-squint condition
\begin{align}
    \sum_{\ell < \Nrf} \!\! \lambda_\ell(\cB) = \|\balpha\| + o(r) \label{upaSquintFree}
\end{align}
holds if and only if $p\geq 1$.
The minimum number of RF chains that asymptotically attain squint-free performance is thus %(see Fig. \ref{fig:projection})
\begin{align}
 \!\!\!   \alpha^{\sf up} & = |\alpha_x|+|\alpha_y| \\    
    & =\frac{W}{\fc} \cdot\frac{L_x |\cos\theta| + L_y |\sin\theta|}{\lambda}\sin \phi , \label{USD}
    % & =\frac{W\sin\phi}{c}\big( \text{array length projected on }(\cos\theta, \sin\theta) \big) 
\end{align}
where $L_x |\cos\theta| + L_y |\sin\theta|$ is the projection of the UPA onto a
line parallel to $\begin{bmatrix}\cos\theta & \sin\theta\end{bmatrix}^\top$ (see Fig. \ref{fig:projection}). In conjunction with $\sin \phi$, this projects the array onto the direction of beamforming as illustrated in Fig. \ref{fig:projectedAperture}.

This result can be readily generalized to non-UPA planar topologies.

\subsection{Beamspace Architecture}

A beamspace architecture analogous to the ULA one in
Sec. \ref{beamSpaceLinear} can be constructed by choosing the $\ell$th column of $\bW_{\sf a}$ as the MRT beamformer $\ba(f_\ell)$ for relative frequency
\begin{align}
    f_\ell = \frac{\ell-\frac{\Nrf-1}{2}}{\alpha^{\sf up}} W.
\end{align}
However, as the columns of $\bW_{\sf a}$ are no longer orthonormal, the proof set forth for linear arrays does not carry over to show that (\ref{USD}) holds for planar arrays. Results in the next section support this, yet the proof is still an open issue.

\begin{table*}
\centering
\caption{Normalized Average Beamforming Gain (Worst Value over Azimuth Angle) for Various Setups} \vspace*{-2mm}
\begin{tabular}{c c c c c c c c c c c c} 
\toprule
 & \# additional & \multicolumn{10}{c}{Bandwidth [GHz]} \\
 & RF chains & 1 & 2 & 3 & 4 & 5 & 10 & 15 & 20 & 25 & 30\\
\midrule 
\multirow{3}{*}{Optimal} & 0 & 0.9877   & 0.9523   & 0.8984   & 0.8324   & 0.9776   & 0.9638   & 0.9680   & 0.9720   & 0.9737   & 0.9749\\
& 1 & 1.0000   & 0.9993   & 0.9967   & 0.9933   & 0.9991   & 0.9961   & 0.9959    & 0.9961   & 0.9960   & 0.9959
 \\
& 2 &  1.0000 &	0.9999 & 0.9999 &	0.9998 &	0.9998 &	0.9996 &	0.9996 &	0.9995 &	0.9994 &	0.9993\\
\midrule
\multirow{3}{*}{Beamspace} & 0 &  0.9876   & 0.9518   & 0.8963   & 0.8266   & 0.9089   & 0.9308   & 0.9431   & 0.9510   & 0.9564   & 0.9603\\
& 1 & 0.8170   & 0.8349   & 0.8600   & 0.8868   & 0.9264   & 0.9405   & 0.9492    & 0.9551   & 0.9596   & 0.9631\\
& 2 &     0.9952   & 0.9819   & 0.9632   & 0.9433   & 0.9531   & 0.9607   & 0.9658   & 0.9695   & 0.9722   & 0.9744\\
\bottomrule
\end{tabular}
\label{nonasymptoticTable}
\vspace*{-4mm}
\end{table*}

\begin{figure*}
    \centering
    \subfloat[Fully-connected]
    {
        \includegraphics[width=0.32\linewidth]{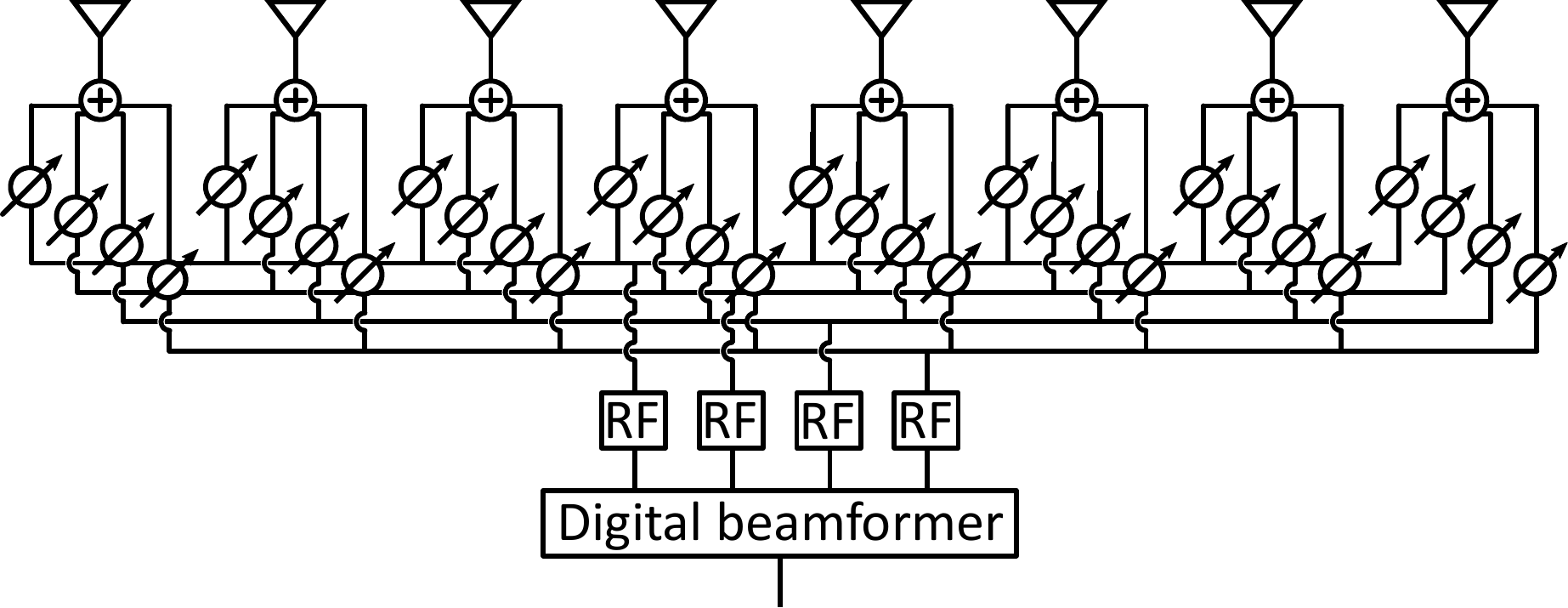}
    }
    \subfloat[Hybridly-connected]
    {
        \includegraphics[width=0.295\linewidth]{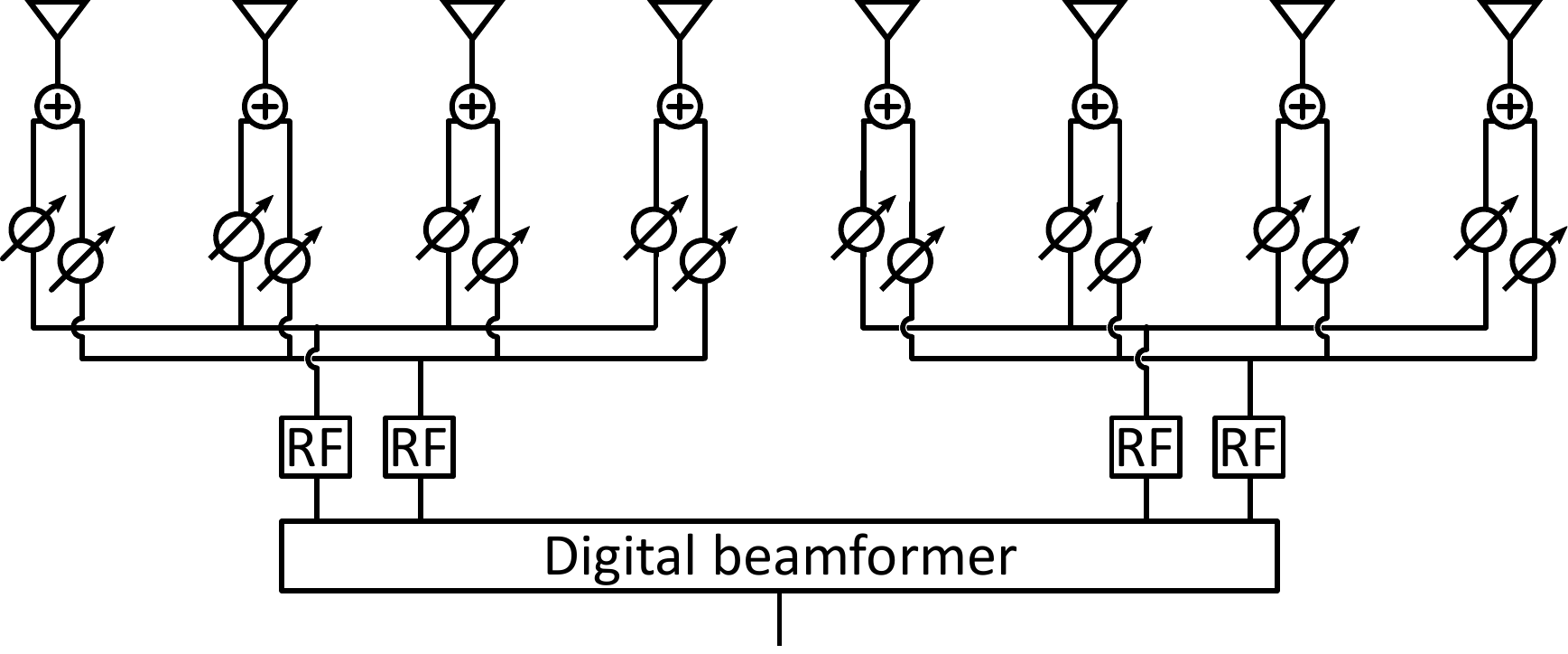}
    }
    \subfloat[Partially-connected]
    {
        \includegraphics[width=0.29\linewidth]{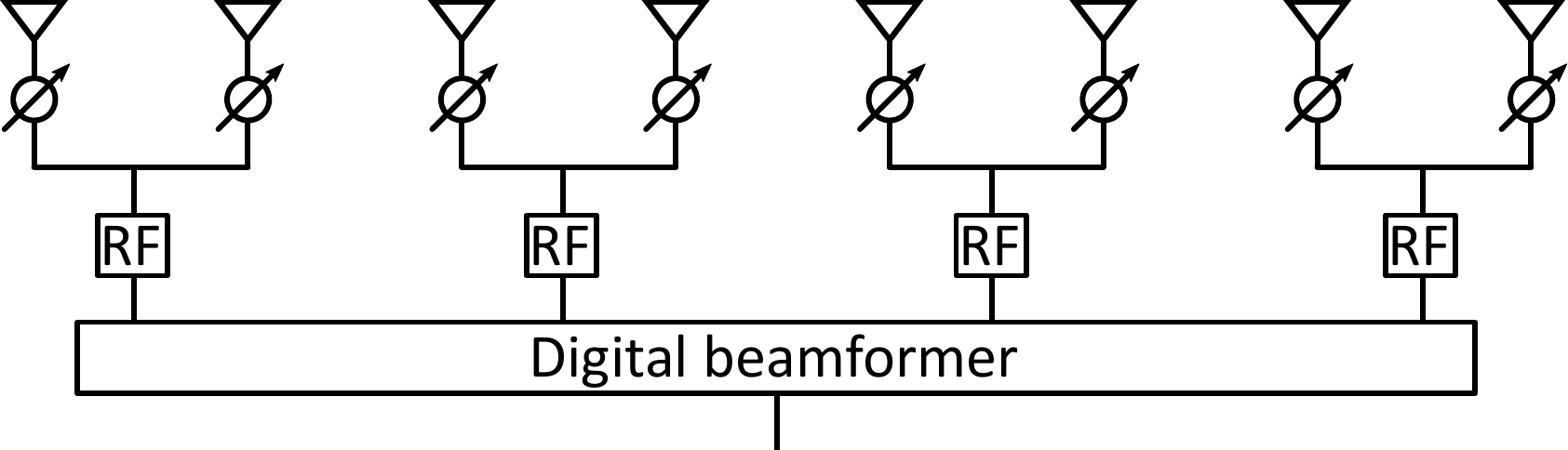}
    }
    \caption{Hybrid architectures: fully-connected, hybridly-connected, and partially-connected.}
    \label{fig:hybridlyConnected}
    \vspace*{-4mm}
\end{figure*}

\section{Numerical Results}
\label{sec:numericalResults}

To further support the relevance of the asymptotic analysis to values of interest, say for 6G or advanced WiFi systems, this section presents additional results. %Precisely, the average beamforming gain in \eqref{gainEigenvalue} is computed for a variety of parameter choices.

A $128\times128$ UPA with half-wavelength spacing is considered, which would occupy only 6.4 cm $\times$ 6.4 cm at 300 GHz.
% At terahertz, such as 300 GHz, $128\times128$ UPA with half-wavelength spacing occupies only 6.4 cm $\times$ 6.4 cm.
% which is much smaller than commercially available antenna arrays for satellite communications; Starlink's phased array for ground station has a footprint of 51.3 cm $\times$ 30.3 cm \cite{starlink2022}. 
% That said, simulating with this large array is prohibitive when it comes to eigenvalue computations. We circumvent this issue by downsampling the array by a factor of four in both directions. The resultant array is then 64$\times$64.
From \eqref{alphaPlanar}, $\balpha$ depends on the product $W\sin\phi$.
%For instance, $W = 30$ GHz with $\phi = 30^\circ$ is basically identical to $W=15$ GHz and $\phi = 90^\circ$.
Without loss of generality we can thus fix $\phi = 90^\circ$ and sweep only $W$.
Table \ref{nonasymptoticTable} lists the average beamforming gain normalized by its squint-free counterpart, $\frac{\gavg}{N}$, for the optimal and beamspace architectures; precisely, the listed values are the worst normalized gains over
$\theta \in \{0^\circ,1^\circ,\ldots,359^\circ\}$. The number of additional RF chains in the table is $\Nrf - \lceil\alpha^{\sf up}\rceil$.
For the optimal architecture, $\lceil\alpha^{\sf up}\rceil +1 $ RF chains ensure over $99\%$ of the squint-free gain over the entire range and, for strong squints, $\lceil\alpha^{\sf up}\rceil +1 $ RF chains suffice to perform almost as impressively. 
%\heedong{With our simulation parameters, we have
%\begin{align}
%    \alpha^{\sf up} = \frac{W}{\fc}\cdot \frac{L_x|\cos\theta|+L_y|\sin\theta|}{\lambda}\in \frac{64 W}{\fc}[1,\sqrt{2}].
%\end{align}
%So for very small bandwidth, say $W= 1,2$ GHz, $\lceil\alpha^{\sf up}\rceil = 1$ is the scenario considered Sec. \ref{sec:beamSquintEffect}. As $\alpha < \alpha_{\sf 3dB} = 0.886$, we're in the small-squint regime. But for $W=3,4$ GHz, there is some $\theta$ where $\alpha^{\sf up}$ is quite large but not larger than 1, i.e., $\alpha_{\sf 3dB}<\Nrf<1$. Perhaps, presenting worst orientation only is not the best way, but I cannot think of better yet simple way. 
%}
As of the beamspace architecture, it exhibits a remarkably small performance deficit relative to the optimal one---a deficit that can be overcome with
only one or two extra RF chains. Given its simplicity, this makes it a decidedly attractive alternative.
%\heedong{Hmm, readers may ask why beam-space architecture performs so poor for small bandwidth and one additional RF chain? It is also true for ULAs. With $\alpha \rightarrow \infty$, \eqref{beamspaceDerivationContinued} becomes
%\begin{align}
%    &\frac{1}{\alpha}\sum_{\ell<\Nrf}\int_{\ell-\frac{\Nrf-1}{2}-\frac{\alpha}{2}}^{\ell-\frac{\Nrf-1}{2}+\frac{\alpha}{2}}\bigg(\frac{\sin \pi t}{\pi t}\bigg)^{\!2} dt\\
%    &= \sum_{\ell<\Nrf} \bigg(\frac{\sin \pi \big(\ell-\frac{\Nrf-1}{2}\big)}{\pi \big(\ell-\frac{\Nrf-1}{2}\big)}\bigg)^{\!2}\\
%    &= \begin{cases}
%    1 & \Nrf = \text{odd}\\
%    \frac{\pi^2}{8}\big(\frac{1}{1^2}+\frac{1}{3^2}+\ldots+ \frac{1}{(\Nrf-1)^2}\big) & \Nrf = \text{even}
%    \end{cases}.
%\end{align}
%For $\Nrf=2$, it is $\frac{\pi^2}{8} \approx 0.81$ (interestingly, it becomes 1 as $\Nrf\rightarrow\infty$). Essentially, the problem is dividing $W$ into subbands spaced by $\frac{W}{\alpha}$, which is very large when $\alpha\ll 1$. This problem can be easily resolved (numerically verified) by replacing \eqref{beamSpaceFrequency} with
%\begin{align}
%    f_\ell =\frac{\ell-\frac{\Nrf}{2}}{\Nrf} W,
%\end{align}
%which is much more sensible from the ``intuition'' of the beam-space architecture.
%But, in analysis-wise, it makes the analog beamformer not column-orthogonal...
%}

\section{Extensions}
\label{sec:extensions}

%\angel{What's the point of the section? Review of existing results? Description of the beam squint phenomenon? It does seem to break the flow. And the notion of "partially-connected analog network" appears out of the blue, without having been motivated or introduced} \heedong{I prepared this section in response to the comment ``there is a lot of work on hybrid precoding with larger arrays. The usual approach would be to have a subarray architecture with the subarrays involving frequency flat beamforming and frequency selective across the subarrays.'' from Prof. Heath. But it seems that I did a poor job. Let me elaborate.}

While the formulation hitherto has considered a general hybrid array, fully-connected as per the illustration in Fig. \ref{fig:hybridlyConnected},
this section turns the attention to the more restrictive hybridly-connected and partially-connected structures, and further to receivers equipped themselves with an array.

\begin{figure*}
    \centering
    \subfloat[$\alpha = 4$]
    {
        \includegraphics[width=0.45\linewidth]{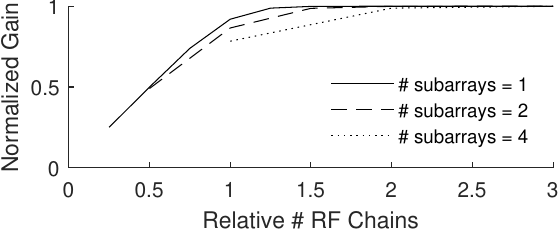}
    }
    \subfloat[$\alpha = 16$]
    {
        \includegraphics[width=0.45\linewidth]{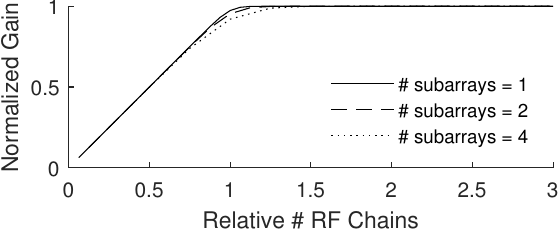}
    }
    \caption{Normalized beamforming gain $\frac{\gavg}{N}$ for a 128-element ULA. The analog beamformer maximizing the average beamforming gain is used. The relative number of RF chains for the horizontal axis is the number of RF chains normalized by the number of required RF chains, that is, $\frac{\Nrf}{\alpha}$.}
    \label{hybridlyBehavior}
    \vspace*{-4mm}
\end{figure*}

\subsection{Hybridly-Connected Architecture}

% \angel{Need to introduce these hybridly-connected architectures, perhaps with a drawing?} \heedong{I'm not perfectly satisfied with Fig. \ref{fig:hybridlyConnected}... Any suggestions?}

%Our central result extends straightforwardly to hybridly-connected architecture \cite{zhang2017hybridly}, where
In a hybridly-connected architecture (see Fig. \ref{fig:hybridlyConnected}), the array is partitioned into subarrays and only a subset of the RF chains is connected to each subarray \cite{zhang2017hybridly}.
With proper antenna indexing, the analog beamforming matrix exhibits a block diagonal structure, namely
\begin{align}
    \bW_{\sf a} = \blkdiag(\bW_{{\sf a},0},\ldots,\bW_{{\sf a},M-1}) \label{blockDiagonal}
\end{align}
with $\bW_{{\sf a},m}$ the analog beamformer for the $m$th subarray and $M$ the number of subarrays. Plugging \eqref{blockDiagonal} into \eqref{gainDefinition} gives
\begin{align}
    g(f)&=\Big\|\ba^*(f)\blkdiag\big(\bW_{{\sf a},0}(\bW_{{\sf a},0}^*\bW_{{\sf a},0})^{-\frac{1}{2}},\nonumber\\
    &\qquad\quad \ldots,\bW_{{\sf a},M-1}(\bW_{{\sf a},M-1}^*\bW_{{\sf a},M-1})^{-\frac{1}{2}}\big)\Big\|^2\\
    &=\sum_m g_m(f) \label{gainHybridlyConnected}
\end{align}
where 
\begin{align}
    g_m(f) = \big\|\ba_m^*(f)\bW_{{\sf a},m}(\bW_{{\sf a},m}^*\bW_{{\sf a},m})^{-\frac{1}{2}}\big\|^2
\end{align}
given $\ba_m(f)$ as the channel for the $m$th subarray. Therefore,
\begin{align}
    \gavg = \sum_m g_{{\sf avg},m} \label{averageGainHybridlyConnected}
\end{align}
where
\begin{equation}
    g_{{\sf avg},m} = \frac{1}{W}\int_{-\frac{W}{2}}^{\frac{W}{2}} g_m(f) df.
\end{equation}
The behavior of a hybridly-connected array can thus be broken down into that of the subarrays, and the squint-free criterion is $g_{{\sf avg},m}$ approximately equaling the number of elements in the $m$th subarray for every $m$.

Let us first consider an $N$-element ULA composed of $M$ subarrays in the asymptotic regime. % in Sec. \ref{sec:linearArrays}.
Each subarray is connected to $\frac{\Nrf}{M}$ RF chains. From Sec. \ref{sec:linearArrays}, squint-free performance requires
\begin{align}
    \frac{\Nrf}{M} \geq \frac{\alpha}{M} \Leftrightarrow \Nrf \geq \alpha,
\end{align}
as in the fully-connected architecture. This is remarkable given the $M$-fold savings in number of phase shifters, yet the asymptotic regime is now determined by $\frac{\alpha}{M}\rightarrow\infty$, whose convergence rate is slower (see Fig. \ref{hybridlyBehavior}).

The situation changes with planar arrays. Let us consider an $N_x \times N_y$ UPA composed of $M_x\times M_y$ subarrays. The asymptotic regime of interest is the one considered in Sec. \ref{sec:planarArrays} with fixed $M_x$ and $M_y$. Each subarray is then connected to $\frac{\Nrf}{M_xM_y}$ RF chains. Attaining squint-free performance requires
\begin{align}
    \frac{\Nrf}{M_xM_y} \geq \frac{|\alpha_x|}{M_x} + \frac{|\alpha_y|}{M_y}\Leftrightarrow \Nrf \geq M_y|\alpha_x| + M_x |\alpha_y|,
\end{align}
which is at least $\min(M_x,M_y)$-times larger than the number of RF chains required by the fully-connected case.

\subsection{Partially-Connected Architecture}

With a single RF chain per subarray, a partially-connected array is the simplest instance of a hybridly-connected architecture (see Fig. \ref{fig:hybridlyConnected}).
% The analog beamformer \eqref{blockDiagonal} can be written as
% \begin{align}
%     \bW_{\sf a} = \blkdiag(\bw_{{\sf a},0},\ldots,\bw_{{\sf a},M-1}),
% \end{align}
% where $\bw_{{\sf a},m}$ is the analog beamformer for the $m$th subarray.
% Recalling \eqref{3dBbandwidthDecreasing}, one can the intensity of beam squint increases when array footprint increases. This observation naturally motivates splitting the array into multiple subarrays, which is called partially-connected architecture.
% Plugging \eqref{blockDiagonal} into \eqref{gainDefinition} gives
% \begin{align}
%     g(f)&=\big\|\ba^*(f)\blkdiag(\bw_{{\sf a},0},\ldots,\bw_{{\sf a},\Nrf-1})\nonumber\\
%     &\qquad\cdot \diag(\|\bw_{{\sf a},0}\|^{-1},\ldots,\|\bw_{{\sf a},\Nrf-1}\|^{-1})\big\|^2\\
%     &=\sum_n g_n(f) \label{gainPartiallyConnected}
% \end{align}
% where $g_n(f) \equiv \frac{|\ba^*_n(f)\bw_{{\sf a},n}|^2}{\|\bw_{{\sf a},n}\|^2}$ given $\ba_n(f)$ the normalized channel for the $n$th subarray.
As per \eqref{averageGainHybridlyConnected}, the behavior of a hybrid array with a partially-connected analog network can be decomposed into that of the subarrays. %Although it is possible to satisfactorily alleviate the beam squint (recall Sec. \ref{sec:beamSquintEffect}), one cannot perfectly eradicate it with this architecture. \angel{Not clear which aspect of Sec. \ref{sec:beamSquintEffect} you're referring to...} \heedong{When the number of subarray grows, $\frac{\alpha}{M}$ shrinks and it becomes less than $\alpha_{\sf 3dB}$ at some point.}
% We shall defer the in-depth discussion on hybrid array architectures till Sec. \ref{sec:extensions}. 

% Partially-connected architectures are often preferred since it can greatly reduce the number of required phase shifters. We shall resume the discussion on this architecture.

%The result \angel{which one?} \heedong{Oh, I was being sloppy. 3dB bandwidth is the bandwidth whereby the worst beamforming gain is the half of the squint-free gain. To compare the analog arrays and fully-connected hybrid arrays in terms of average beamforming gain, we need to compute the average beamforming gain for analog arrays (it is already computed for fully-connected hybrid arrays).} in Sec. \ref{sec:beamSquintEffect} is on the worst-case beamforming gain. To compare it with the fully-connected architecture, we extend the analysis to the average beamforming gain. 
Consider an $N$-element ULA composed of $M=\Nrf$ subarrays with identical subarray beamformer $\bw_{\sf a}\in\bbC^{\frac{N}{M}}$. As per \eqref{averageGainHybridlyConnected},
% \footnote{Since all subarrays are assumed to be identical, we choose the zeroth subarray.}
$    \gavg = Mg_{{\sf avg},0} \label{gainPartiallyConnectedIdentical}$.
Two subarray beamformers are considered, the MRT beamformer
%in Sec. \ref{sec:beamSquintEffect}
and the optimal beamformer,
%in Sec. \ref{optimalBeamformer}. Both can be
both implementable with a single RF chain and $M$ delay lines (see App. \ref{partiallyConnectedProperty}).

For the MRT beamformer,
\begin{align}
    \gavg = \frac{M}{W}\int_{-\frac{W}{2}}^{\frac{W}{2}} F_{\frac{N}{M}} \! \bigg(\frac{2\pi d_x u_x}{c}f\bigg) df
\end{align}
and thus
\begin{align}
    \frac{\gavg}{N} &\rightarrow \frac{1}{W} \int_{-\frac{W}{2}}^{\frac{W}{2}} \bigg(\frac{\sin(\frac{\pi L_x u_x}{M c}f)}{\frac{\pi L_x u_x}{M c}f}\bigg)^{\! 2} df\\
    &=\frac{M}{\pi \alpha} \int_{-\frac{\pi\alpha}{2M}}^{\frac{\pi\alpha}{2M}} \bigg(\frac{\sin t}{t}\bigg)^{\! 2} dt\\
    &=\frac{2M}{\pi \alpha} \bigg(\operatorname{Si} \! \bigg(\frac{\pi\alpha}{M}\bigg)-\frac{\sin^2 \frac{\pi\alpha}{2M}}{\frac{\pi\alpha}{2M}}\bigg) \label{gainPartialMrt}
\end{align}
as the array densifies; here, $\operatorname{Si}(\cdot)$ is the sine integral.

Let us now consider the optimal beamformer in terms of average beamforming gain. 
% With such a beamformer, we have
% \begin{align}
%     g_{{\sf avg},0} = \lambda_0(\bB_{\sf subarray}),
% \end{align}
% where $\bB_{\sf subarray} \in \bbR^{M\times M}$ is a matrix with entries 
% \begin{align}
%     [\bB_{\sf subarray}]_{n',n} =\frac{\sin \! \big( \pi \frac{W u_xd_x}{c} (n-n')\big)}{\pi \frac{W u_xd_x}{c} (n-n')}.
% \end{align}
% This results in
% \begin{align}
%     \frac{\gavg}{N} \leq \frac{\lambda_0(\bB_{\sf subarray})}{M},
% \end{align}
% which is essentially \cite[Prop. 2]{park2017dynamic}.
As array densifies, as per \eqref{eigenvalueRelation},
\begin{align}
    \frac{\gavg}{N} \rightarrow  \frac{\lambda_0(\cB_{\alpha/M})}{\alpha/M} \label{partiallyConnectedLimit}.
\end{align}

\begin{figure}
    \centering
    \includegraphics[width=0.9\linewidth]{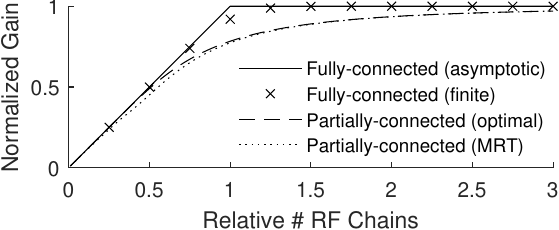}
    \caption{Comparison of array architectures. For the fully-connected architecture, $\alpha\rightarrow \infty$ and $\alpha=4$ are considered. The relative number of RF chains for the horizontal axis is the number of RF chains normalized by the number of required RF chains, that is, $\frac{\Nrf}{\alpha}$. This equals $\frac{M}{\alpha}$ for partially-connected architectures.}
    \label{architectureComparison}
    \vspace*{-4mm}
\end{figure}

%\begin{figure}
%    \centering
%    \includegraphics[width=0.9\linewidth]{Images/architectureComparisonDb.eps}
%    \caption{Comparison of array architectures in dB scale.}
%    \label{architectureComparisonDb}
%\end{figure}

Fig. \ref{architectureComparison} compares the normalized average beamforming gain for fully- and partially-connected architectures. Recalling \eqref{ulaSquintFree}, the asymptotic result $\min(p,1)$ is plotted for the fully-connected architecture while \eqref{gainPartialMrt} and \eqref{partiallyConnectedLimit} are plotted for the partially-connected architecture.

% From the positivity of eigenvalues \cite{slepian1961prolate}, we cannot achieve the squint-free performance
% \begin{align}
%     \lambda_0(\cB_{\alpha/\Nrf}) = \frac{\alpha}{\Nrf} - \sum_{n>0} \lambda_0(\cB_{\alpha/\Nrf}) < \frac{\alpha}{\Nrf}
% \end{align}
% in a strict mathematical sense. That said, it is virtually achievable with reasonable number of RF chains.
Although the squint cannot be completely eradicated with a partially-connected architecture, it is rather enticing because of the $M$-fold save in the number of phase shifters.
%For instance, with $M = \alpha, 2\alpha, 3\alpha$, a partially-connected architecture incurs respective losses of only $1.11$, $0.29$, $0.13$ dB, even with the MRT beamformer (see Fig. \ref{architectureComparison}).
For instance, with $M = \alpha, 2\alpha, 3\alpha$, a partially-connected architecture incurs respective losses of only $1.11$, $0.29$, $0.13$ dB, even with simple MRT beamforming; in Fig. \ref{architectureComparison}, these losses can be appreciated in linear scale.

Similar to their hybridly-connected brethren, partially-connected architectures
become less alluring with planar arrays.
To maintain the subarray size, one needs the scalings
\begin{align}
    M_x \propto \alpha_x \qquad M_y \propto \alpha_y ,
\end{align}
whereby the substantial disadvantage of the separable architecture in Sec. \ref{separableArchitecture} is reproduced.
% Let us consider $N_x \times N_y$ UPA composed of $M_x\times M_y$ subarrays where $M_x = \frac{N_x}{N_{{\rm RF},x}}\in\bbZ$ and $M_y = \frac{N_y}{N_{{\rm RF},y}}\in\bbZ$. 
% Repeating the analysis, the normalized average beamforming gain is bounded by
% \begin{align}
%     \frac{ \lambda_0(\cB_{\balpha^{\sf subarray}})}{\|\balpha^{\sf subarray}\|} \label{partiallyConnectedUpa}
% \end{align}
% where $\balpha^{\sf subarray} \equiv \begin{bmatrix} \frac{\alpha_x}{N_{{\rm RF},x}} & \frac{\alpha_y}{N_{{\rm RF},y}} \end{bmatrix}$. Consider a scenario where $\alpha_x$ and $\alpha_y$ grow with a fixed ratio.
% To maintain $\balpha^{\sf subarray}$, one need to ensure
% % $N_{{\rm RF},x}\propto \alpha_x$ and $N_{{\rm RF},y}\propto \alpha_y$,
% \begin{align}
%     N_{{\rm RF},x}\propto \alpha_x \qquad N_{{\rm RF},y}\propto \alpha_y
% \end{align}
% sharing the substantial disadvantage of the separable architecture in Sec. \ref{separableArchitecture}.
This shortcoming can be alleviated by a use of dynamic subarrays \cite{park2017dynamic}, but at the expense of additional hardware complexity.

\subsection{Multiantenna Receiver}

The separable property of far-field LOS multiantenna channels enables the extension to such a setting too.
Positing an $\Nt$-element transmitter and an $\Nr$-element receiver, the LOS channel can be expressed as \cite[Sec. II-A]{do2023parabolic}
\begin{align}
    \bH(f) = \ba_{\rm r}(f)\ba^*_{\rm t}(f)
\end{align}
where $\ba_{\rm t}(f)\in \bbC^{\Nt}$ is the normalized channel between the transmit array and the receive array center, and reciprocally for $\ba_{\rm r}(f)\in \bbC^{\Nr}$. The squint-free condition is $\gavg \approx \Nt\Nr$.

Let us denote the analog and digital beamformers by $\bW_{{\sf a}, {\rm t}}$ and $\bw_{{\sf d}, {\rm t}}(f)$ for the transmitter, and $\bW_{{\sf a}, {\rm r}}$ and $\bw_{{\sf d}, {\rm r}}(f)$ for the receiver. An analogous signal model to that in Sec. \ref{signalModel} is
\begin{align}
    y(f) &= \bw_{{\sf d}, {\rm r}}^*(f)\bW_{{\sf a},{\rm r}}^*\big(\bH(f)\bW_{{\sf a}, {\rm t}}\bw_{{\sf d}, {\rm t}}(f) s(f) \!+\! \bv(f)\big)
\end{align}
where each entry of $\bv(f)$ describes a white Gaussian noise process with unit power spectral density. Letting $s(f)$ have unit power, the power constraint becomes
\begin{align}
    \frac{1}{W}\int_{-\frac{W}{2}}^{\frac{W}{2}} \|\bW_{{\sf a}, {\rm t}}\bw_{{\sf d}, {\rm t}}(f)\|^2 df = \SNR.
\end{align}
Also, without loss of generality, we let $\|\bW_{{\sf a},{\rm r}}\bw_{{\sf d}, {\rm r}}(f)\| = 1$.

Reiterating the simplifications made in Sec. \ref{signalModel} gives
\begin{align}
    y(f)= \sqrt{g(f)p(f)}s(f) + \bW_{{\sf a},{\rm r}}^*\bw_{{\sf d}, {\rm r}}^*(f) \bv(f)
\end{align}
where
\begin{align}
    g(f) &= \|\bw_{{\sf d}, {\rm r}}^*(f)\bW_{{\sf a},{\rm r}}^*\bH(f)\bW_{{\sf a}, {\rm t}}\bw_{{\sf d}, {\rm t}}(f)\|^2\\
    &= |\ba^*_{\rm t}(f)\bW_{{\sf a}, {\rm t}}\bw_{{\sf d}, {\rm t}}(f)|^2 \cdot |\ba_{\rm r}^*(f)\bW_{{\sf a},{\rm r}}\bw_{{\sf d},{\rm r}}(f)|^2\\
    &= g_{\rm t}(f) g_{\rm r}(f)
\end{align}
with
\begin{align}
g_{\rm t}(f) & = |\ba^*_{\rm t}(f)\bW_{{\sf a}, {\rm t}}\bw_{{\sf d}, {\rm t}}(f)|^2 \\
g_{\rm r}(f) & = |\ba_{\rm r}^*(f)\bW_{{\sf a},{\rm r}}\bw_{{\sf d},{\rm r}}(f)|^2.
\end{align}
The effective noise $\bW_{{\sf a},{\rm r}}^*\bw_{{\sf d}, {\rm r}}^*(f) \bv(f)$
is a white Gaussian process with unit power spectral density.

Recalling the definition of average beamforming gain in (\ref{gavgdef}),
%Denoting $\gavg \equiv \frac{1}{W}\int_{-\frac{W}{2}}^{\frac{W}{2}} g(f) df$, we have
it is shown in App. \ref{separableProof} that
\begin{align}
    \frac{g_{{\sf avg},{\rm t}}}{\Nt} + \frac{g_{{\sf avg},{\rm r}}}{\Nt} - 1 \leq \frac{\gavg}{\Nt\Nr} \leq \min \! \bigg(\frac{g_{{\sf avg},{\rm t}}}{\Nt}, \frac{g_{{\sf avg},{\rm r}}}{\Nt}\bigg).
\end{align}
From this result, plus $g_{{\sf avg},{\rm t}}\leq \Nt$ and $g_{{\sf avg},{\rm r}}\leq \Nr$, it follows that
$\frac{g_{{\sf avg},{\rm t}}}{\Nt} = \frac{g_{{\sf avg},{\rm r}}}{\Nr} =1$
% \begin{equation}
% \frac{g_{{\sf avg},{\rm t}}}{\Nt} = \frac{g_{{\sf avg},{\rm r}}}{\Nr} =1
% \end{equation}
is a necessary and sufficient condition for $\frac{\gavg}{\Nt\Nr}=1$.
This implies that both $g_{{\sf avg},{\rm t}} \approx \Nt$ and $g_{{\sf avg},{\rm r}} \approx \Nr$ are needed to ensure $\gavg \approx \Nt\Nr$, and vice versa, whereby the multiantenna problem is seen to decouple into two problems that have already been addressed.

\begin{figure}
    \centering
    \includegraphics[width=0.7\linewidth]{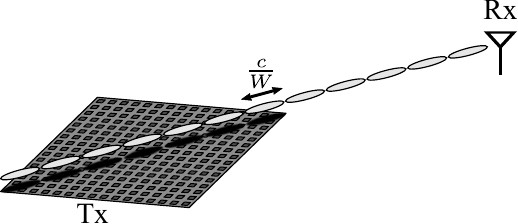}
    \caption{A train of symbols, each of length $\frac{c}{W}$, traveling over the array along the worst possible direction in terms of the symbol delay across element.}
    \label{fig:intuition}
    \vspace*{-4mm}
\end{figure}

\section{Conclusion}
\label{sec:conclusion}
The main takeaway point of this work is that hybrid arrays can operate free of beam squint, just like digital arrays, only with a much smaller number of RF chains.
Although derived asymptotically in the bandwidth-aperture product, this result applies for virtually any value of that product.

For a hybrid linear array designed to beamform in any direction, the needed number of RF chains is $\frac{1}{c} W L$. In contrast, a digital array requires a chain per antenna, which for half-wavelength element spacing amounts to $\frac{L}{\lambda/2} = \frac{2}{c} \fc L$. The contrast between the two expressions evidences something fundamental, namely that it is the bandwidth rather than the carrier frequency that matters to the beam squint. Indeed, $\frac{c}{W}$ is roughly the spatial length of a symbol, and squint arises once symbols cease to be much longer than the array (see Fig. \ref{fig:intuition}).
The value $\frac{1}{c} W L$ equals the number of symbols that fit on the array along the worst possible direction, and using that many RF chains ensures one digital sample per symbol.

For planar arrays,
the ratio between the RF chains required by digital and hybrid arrays is even more pronounced, as planar arrays can use space more efficiently \cite{nguyen2022beam}.

%\heedong{Here is the calculation. The amount of reduction equals
%\begin{align}
%    \frac{N}{\max_{\bu}\alpha^{\sf up}}&=\frac{N}{\frac{W}{\fc}\max_{\bu}\frac{L_x|u_x|+L_y|u_y|}{\lambda}}\\
%    &=\frac{\fc}{W}\cdot \frac{N\lambda}{\sqrt{L_x^2+L_y^2}} \label{doubleCheck}\\
%    &=\frac{2\fc}{W}\cdot\frac{N_xN_y}{\sqrt{N_x^2+N_y^2}}\\
%    &\leq \sqrt{2N}\frac{\fc}{W}
%\end{align}
%The last step follows from AM-GM inequality. Fixing the total number of antennas, square arrangement is the best, where the diagonal array dimension is minimized.}
%\heedong{This result basically says that planar arrangement is much better than linear arrangement as $L_x\times L_y$ UPA performs similar to $\sqrt{L_x^2+L_y^2}$ ULA. Similar conclusion is made in \cite{nguyen2022beam} for analog arrays.}
%\angel{Hmm. With a digital array the number of chains equals $\frac{L_x}{\lambda/2} \frac{L_y}{\lambda/2} = \frac{4 L_x L_y \fc^2}{c^2}$, with a hybrid array it's $\frac{W \sqrt{l_x^2 + l_y^2}}{c}$, yes?} \heedong{Right. I double-checked the calculation. It again gives the ratio \eqref{doubleCheck}. The additional gain depending on $N$ follows from departing from separable design. With separable design, for the worst beam direction, we need $\frac{1}{2}\cdot\frac{WL_x}{c}\cdot\frac{WL_y}{c}$ RF chains, and the ratio does not depend on the array size.}

Notably, the above points remain valid for suboptimum beamspace architectures, and in the case of linear arrays even for hybridly-connected structures, with the concomitant reduction in number of phase shifters; for hybridly-connected planar arrays, the reduction in phase shifters entails a tradeoff with the increase in RF chains.

%ON HYBRIDLY-CONNECTED:
%For linear arrays, the number of phase shifters can be greatly reduced with a reasonable number of additional RF chains. However, for planar arrays, this trade-off becomes unfavorable; we need much more additional RF chains when reducing the number of phase shifters.

%The analysis in this paper invites research on the
%A number of ... loom in the horizon
Further research is required to determine whether similar behaviors are encountered
in multipath channels, with intelligent surfaces \cite{park2022beam},
in hybrid architectures with true-time-delay \cite{zhai2020thzprism, dai2022delay, steinweg2022hybrid, jain2023mmflexible}, or in near-field situations \cite{myers2021infocus}.

%Moving forward, a natural continuation of this work would be to explore the extension of hybrid arrays in multipath environments.
% Additionally, the significance of multibeam satellites in contemporary communications necessitates the exploration of multi-user setups, with related works \cite{zheng2012generic, vazquez2016precoding, you2020massive} providing valuable insight.
%The present work is restricted to traditional array antennas, and therefore, investigating the suitability of hybrid arrays for intelligent reflecting surfaces or lenses \cite{park2022beam} would require a new analysis, and analogous results may not apply.
%Moreover, incorporating true-time-delay in hybrid array architecture \cite{zhai2020thzprism, dai2022delay, steinweg2022hybrid, jain2023mmflexible} would also be an intriguing direction for future research.
%\heedong{EXTENSION TO NEAR-FIELD SCENARIOS IS ALSO AN INTERESTING DIRECTION! \cite{myers2021infocus}.
%UNFORTUNATELY, NEAR-FIELD MIMO CHANNEL IS NO LONGER SEPARABLE SO THERE'S LITTLE HOPE THAT MISO/SIMO CASE NATURALLY GENERALIZES TO MIMO CASE \cite{do2023parabolic}.}

\appendices

\section{}
\label{3dBbandwidthProof}
% For brevity, let us define $t= \frac{N}{2} \cdot \frac{\pi  d_x u_x}{c} W_{\sf 3dB}$. 
% Then, \eqref{3dBbandwidth} can be rewritten as
% \begin{align}
%     \frac{\sin t}{N\sin \frac{t}{N}} = \frac{1}{\sqrt{2}}. \label{3dBbandwidthSimple}
% \end{align}
As the smallest solution to \eqref{alpha3db} lies within the main lobe, it suffices to consider $0 < \alpha_{\sf 3dB} < 2$ and the problem reduces to
\begin{align}
    \frac{\sin\frac{\pi\alpha_{\sf 3dB}}{2}}{N\sin\frac{\pi\alpha_{\sf 3dB}}{2N}} = \frac{1}{\sqrt{2}}. \label{3dBbandwidthSimple}
\end{align}
Manipulating the left-hand side into
\begin{align}
    \frac{\sin \frac{\pi \alpha_{\sf 3dB}}{2}}{N\sin \frac{\pi \alpha_{\sf 3dB}}{2N}} = \frac{\sin \frac{\pi \alpha_{\sf 3dB}}{2}}{\frac{\pi \alpha_{\sf 3dB}}{2}}\cdot\frac{\frac{\pi \alpha_{\sf 3dB}}{2N}}{\sin \frac{\pi \alpha_{\sf 3dB}}{2N}}
\end{align}
shows that it is decreasing in $N$. Paired with the fact that the left-hand side is decreasing in $\alpha_{\sf 3dB}$, the solution of \eqref{3dBbandwidthSimple} is decreasing with respect to $N$.

% For a complete understanding, it is beneficial to consider the extreme cases: $N=2$ and $N\rightarrow \infty$. Both are straightforward. For $N=2$, \eqref{3dBbandwidthSimple} becomes $\cos\frac{t}{2} = \frac{1}{\sqrt{2}}$,
% i.e., $t = \frac{\pi}{2}$. For $N\rightarrow \infty$, \eqref{3dBbandwidthSimple} is reduced to $\frac{\sin t}{\bar{x}} = \frac{1}{\sqrt{2}}$,
% giving $t \approx 0.443\pi$.

% \section{}
% \label{partiallyConnectedProof}
% Let us assume that all subarrays are exactly identical in topology (only with translation) and phase shifts. Denoting the displacement of $n$th subarray with respect to $0$th subarray by $\overline{\br}_n$, we have
% \begin{align}
%     \ba_n^*(f) = \exp \! \bigg(j2\pi \frac{\fc+f }{c}\bu^\top \overline{\br}_n\bigg) \ba_0^*(f).
% \end{align}
% We also assume $\bw_{{\sf a},0} = \ldots = \bw_{{\sf a},\Nrf-1}$.
% The exponential term in \eqref{translationSubarray} can be compensated by TTD, resulting in the beamforming gain of
% \begin{align}
%     \bigg|\frac{1}{\sqrt{N}}\sum_n \ba_0^*(f)\bw_{{\sf a},n}\bigg|^2 = N|\ba_0^*(f)\bw_{{\sf a},0}|^2,
% \end{align}
% which is identical to \eqref{gainPartiallyConnected}.

\section{}
\label{ulaProof}

Combining $\lambda_\ell(\cB_\alpha)\leq 1$ and $\sum_\ell \lambda_\ell(\cB_\alpha) = \alpha$, we have that
\begin{align}
    \sum_{\ell<\Nrf}\lambda_n(\cB_\alpha) \leq \min(p,1) \alpha. \label{singularValueBoundsCombined}
\end{align}
In turn, for $\epsilon>0$, we also have that
\begin{align}
    \sum_{\ell<\Nrf}\lambda_\ell(\cB_\alpha) &> \epsilon \big|\{\ell:\sigma_\ell >\epsilon, \ell<\Nrf\}\big|\\
    &= \epsilon \min(\alpha+\cO(\log \alpha), \lceil p\alpha \rceil)
\end{align}
where the last equality follows from \eqref{polarizationFormal}. Consequently
\begin{align}
    &\liminf_{\alpha\rightarrow \infty} \frac{\sum_{\ell<\Nrf}\lambda_\ell(\cB_\alpha)}{\alpha}\nonumber\\
    &\qquad \geq \epsilon \liminf_{\alpha\rightarrow \infty} \frac{\min(\alpha+\cO(\log \alpha), \lceil p\alpha \rceil)}{\alpha}\\
    &\qquad = \epsilon \min(p,1).
\end{align}
where, as the existence of the limit is not guaranteed at this point, limit inferior was used.

For $\epsilon$ arbitrarily close to $1$,
\begin{align}
    \liminf_{\alpha\rightarrow \infty} \frac{\sum_{\ell<\Nrf}\lambda_\ell(\cB_\alpha)}{\alpha} = \min(p,1).
\end{align}
From \eqref{singularValueBoundsCombined}, the limit inferior and limit superior coincide, hence the limit does exist and it equals $\min(p,1)$ as desired.

\section{}
\label{beamspaceProof}

From $\int_{-\infty}^{\infty} \big(\frac{\sin \pi t}{\pi t}\big)^2dt=1$, which is straightforward from Parseval's identity,
\begin{align}
    \frac{1}{\alpha}\sum_{\ell<\lceil p\alpha \rceil}\int_{\ell-\frac{\lceil p\alpha \rceil-1}{2}-\frac{\alpha}{2}}^{\ell-\frac{\lceil p\alpha \rceil-1}{2}+\frac{\alpha}{2}}\bigg(\frac{\sin \pi t}{\pi t}\bigg)^{\! 2} dt \leq \frac{\lceil p\alpha \rceil}{\alpha}.
\end{align}
Also, it can be readily shown that
\begin{align}
    &\frac{1}{\alpha}\sum_{\ell<\lceil p\alpha \rceil}\int_{\ell-\frac{\lceil p\alpha \rceil-1}{2}-\frac{\alpha}{2}}^{\ell-\frac{\lceil p\alpha \rceil-1}{2}+\frac{\alpha}{2}}\bigg(\frac{\sin \pi t}{\pi t}\bigg)^{\!2} dt\nonumber\\
    & \qquad\qquad\leq \frac{1}{\alpha}\sum_{\ell}\int_{\ell-\frac{\lceil p\alpha \rceil-1}{2}-\frac{\alpha}{2}}^{\ell-\frac{\lceil p\alpha \rceil-1}{2}+\frac{\alpha}{2}}\bigg(\frac{\sin \pi t}{\pi t}\bigg)^{\! 2} dt\\
    & \qquad\qquad=\frac{1}{\alpha} \cdot \alpha \int_{-\infty}^{\infty} \bigg(\frac{\sin \pi t}{\pi t}\bigg)^{\! 2} dt = 1.
\end{align}
Combining them, we obtain
\begin{align}
    \limsup_{\alpha\rightarrow \infty} \frac{1}{\alpha}\sum_{\ell<\lceil p\alpha \rceil}\int_{\ell-\frac{\lceil p\alpha \rceil-1}{2}-\frac{\alpha}{2}}^{\ell-\frac{\lceil p\alpha \rceil-1}{2}+\frac{\alpha}{2}}\bigg(\frac{\sin \pi t}{\pi t}\bigg)^{\! 2} dt \leq \min(p,1).
\end{align}

\begin{figure*}
\begin{align}
 &    \!\!\!\!\!\!\!\!\!\!\!\! \frac{1}{\alpha}\sum_{\ell<\lceil p\alpha \rceil}\int_{\ell-\frac{\lceil p\alpha \rceil-1}{2}-\frac{\alpha}{2}}^{\ell-\frac{\lceil p\alpha \rceil-1}{2}+\frac{\alpha}{2}}\bigg(\frac{\sin \pi t}{\pi t}\bigg)^{\! 2} dt \nonumber\\
    &\geq \frac{1}{\alpha}\sum_{\ell} \big[0\leq \ell<\lceil p\alpha \rceil\big]\bigg[\ell-\frac{\lceil p\alpha \rceil-1}{2}+\frac{\alpha}{2}>\epsilon\alpha\bigg]\bigg[\ell-\frac{\lceil p\alpha \rceil-1}{2}-\frac{\alpha}{2}<-\epsilon\alpha \bigg] \int_{\ell-\frac{\lceil p\alpha \rceil-1}{2}-\frac{\alpha}{2}}^{\ell-\frac{\lceil p\alpha \rceil-1}{2}+\frac{\alpha}{2}}\bigg(\frac{\sin \pi t}{\pi t}\bigg)^{\!2} dt \label{discardingSomeTerms}\\
    &\geq \frac{1}{\alpha}\sum_{\ell} \big[0\leq \ell<\lceil p\alpha \rceil\big]\bigg[\ell-\frac{\lceil p\alpha \rceil-1}{2}+\frac{\alpha}{2}>\epsilon\alpha\bigg]\bigg[\ell-\frac{\lceil p\alpha \rceil-1}{2}-\frac{\alpha}{2}<-\epsilon\alpha \bigg] \int_{-\epsilon\alpha}^{\epsilon\alpha}\bigg(\frac{\sin \pi t}{\pi t}\bigg)^{\!2} dt \label{discardingInterval}\\
    &= \Bigg(\frac{1}{\alpha}\sum_{\ell} \big[0\leq \ell<\lceil p\alpha \rceil\big]\bigg[\ell-\frac{\lceil p\alpha \rceil-1}{2}+\frac{\alpha}{2}>\epsilon\alpha\bigg]\bigg[\ell-\frac{\lceil p\alpha \rceil-1}{2}-\frac{\alpha}{2}<-\epsilon\alpha \bigg]\Bigg) \int_{-\epsilon\alpha}^{\epsilon\alpha} \bigg(\frac{\sin \pi t}{\pi t}\bigg)^{\! 2} dt \label{factoringOutConstant}\\
    &\rightarrow \min \! \bigg(p,\frac{1+p-\epsilon}{2}\bigg)-\max \! \bigg(0,\frac{-1+p+\epsilon}{2}\bigg) \label{computingLimit}
\end{align}
\hrulefill
\vspace*{-4mm}
\end{figure*}

At the same time, for $\epsilon>0$ we can manipulate \eqref{beamspaceDerivationContinued} as in \eqref{discardingSomeTerms}--\eqref{computingLimit}, where \eqref{discardingSomeTerms} arises by imposing additional constraint on the summation index, \eqref{discardingInterval} by restricting the interval of integration, \eqref{factoringOutConstant} by factoring out the constant term with respect to the summation index, and \eqref{computingLimit} by computing the limit. Hence,
\begin{align}
    &\liminf_{\alpha\rightarrow \infty} \frac{1}{\alpha}\sum_{\ell<\lceil p\alpha \rceil}\int_{\ell-\frac{\lceil p\alpha \rceil-1}{2}-\frac{\alpha}{2}}^{\ell-\frac{\lceil p\alpha \rceil-1}{2}+\frac{\alpha}{2}}\bigg(\frac{\sin \pi t}{\pi t}\bigg)^{\! 2} dt\nonumber\\
    &\geq \min \! \bigg(p,\frac{1+p-\epsilon}{2}\bigg)-\max \! \bigg(0,\frac{-1+p+\epsilon}{2}\bigg).
\end{align}
For an arbitrarily small $\epsilon>0$,
\begin{align}
    &\liminf_{\alpha\rightarrow \infty} \frac{1}{\alpha}\sum_{\ell<\lceil p\alpha \rceil}\int_{\ell-\frac{\lceil p\alpha \rceil-1}{2}-\frac{\alpha}{2}}^{\ell-\frac{\lceil p\alpha \rceil-1}{2}+\frac{\alpha}{2}}\bigg(\frac{\sin \pi t}{\pi t}\bigg)^{\!2} dt\nonumber\\
    &\geq \min \! \bigg(p,\frac{1+p}{2}\bigg)\!-\!\max \! \bigg(0,\frac{-1+p}{2}\bigg)=\min(p,1),
\end{align}
which concludes the proof.

\section{}
\label{separableProof}

This appendix identifies the maximum and minimum of
\begin{align}
    \frac{1}{W}\int_{-\frac{W}{2}}^{\frac{W}{2}} \frac{g_{\rm t}(f)g_{\rm r}(f)}{\Nt\Nr} df 
\end{align}
under the constraints
\begin{align}
    &0\leq g_{\rm t}(f)\leq \Nt && \frac{1}{W}\int_{-\frac{W}{2}}^{\frac{W}{2}} g_{\rm t}(f) df = g_{{\sf avg},{\rm t}} \nonumber\\
    &0 \leq g_{\rm r}(f) \leq \Nr && \frac{1}{W}\int_{-\frac{W}{2}}^{\frac{W}{2}} g_{\rm r}(f) df = g_{{\sf avg},{\rm r}}.
\end{align}
Introducing 
\begin{align}
    f_{\rm t}(t) \equiv \frac{g_{\rm t}\big(W\big(t-\frac{1}{2}\big)\big)}{\Nt} \qquad f_{\rm r}(t) \equiv \frac{g_{\rm r}\big(W\big(t-\frac{1}{2}\big)\big)}{\Nr},
\end{align}
the objective and the constraints become
\begin{align}
    \int_0^1 f_{\rm t}(t)f_{\rm r}(t) df 
\end{align}
and
\begin{align}
    &0\leq f_{\rm t}(t)\leq 1 && \int_0^1 f_{\rm t}(t) df = g_{{\sf avg},{\rm t}} \nonumber\\
    &0 \leq f_{\rm r}(t) \leq 1 && \int_0^1 f_{\rm r}(t) df = g_{{\sf avg},{\rm r}}.
\end{align}

% The result can be easily generalized to arbitrary interval of integration via a change of variables. \heedong{NOTATION ISSUE...} \angel{Which one?} \heedong{$g_{{\sf avg},1}$ and $g_{{\sf avg},1}$ are used in \eqref{averageGainHybridlyConnected}. Replacing "1" and "2" with "t" and "r" would work. \angel{Yes, good solution} I'll work on it later. Also, I missed the justification of letting $W=1$ without loss of generality.}

Commencing with the maximum value,
\begin{align}
    \int_0^1 f_{\rm t}(t)f_{\rm r}(t) dt \leq \int_0^1 f_{\rm t}(t) dt = g_{{\sf avg},{\rm t}}.
\end{align}
Repeating the argument gives 
\begin{align}
    \int_0^1 f_{\rm t}(t)f_{\rm r}(t) dt \leq \int_0^1 f_{\rm r}(t) dt = g_{{\sf avg},{\rm r}}.
\end{align}
Combining the bounds, 
\begin{align}
    \int_0^1 f_{\rm t}(t)f_{\rm r}(t) dt \leq \min\big(g_{{\sf avg},{\rm t}}, g_{{\sf avg},{\rm r}}\big)
\end{align}
This bound is tight in that it can be attained by 
\begin{align}
    f_{\rm t}(t) = [0\leq t \leq g_{{\sf avg},{\rm t}}] \qquad f_{\rm r}(t) = [0\leq t \leq g_{{\sf avg},{\rm r}}].
\end{align}

Turning to the minimum value, it can be obtained from the observation that
\begin{align}
    \int_0^1 (1-f_{\rm t}(t))(1-f_{\rm r}(t)) dt \geq 0,
\end{align}
which is equivalent to
\begin{align}
    \int_0^1 f_{\rm t}(t)f_{\rm r}(t) dt \geq g_{{\sf avg},{\rm t}} + g_{{\sf avg},{\rm r}} - 1.
\end{align}
Together with the nonnegativity, we have that
\begin{align}
    \int_0^1 f_{\rm t}(t)f_{\rm r}(t) dt \geq \max\big(g_{{\sf avg},{\rm t}} + g_{{\sf avg},{\rm r}} - 1, 0\big).
\end{align}
This bound can be attained by
\begin{align}
    f_{\rm t}(t) \!=\! [0\!\leq\! t \!\leq\! g_{{\sf avg},{\rm t}}] \qquad f_{\rm r}(t) \!=\! [1\!-\!g_{{\sf avg},{\rm r}} \!\leq\! t \!\leq\! 1].
\end{align}

\section{}
\label{simplificationProof}

Let us consider a spectral decomposition of $\tilde{B}_{\balpha}$,
\begin{align}
    \tilde{B}_{\balpha}(\tilde{r}_x',\tilde{r}_x) = \sum_{\ell} \lambda_\ell u_\ell(\tilde{r}_x') \overline{u_\ell(r_x)}, \label{spectralDecomposition}
\end{align}
where $\{u_\ell\}$ is a set of orthonormal functions. Multiplying both sides with 
\begin{align}
    \frac{\big[\tilde{\br} \in \bR A\big]}{\big(\int \big[\tilde{\br} \in \bR A\big] d\tilde{r}_y\big)^{\frac{1}{2}}} \cdot \frac{\big[ \tilde{\br}' \in \bR A\big]}{\big(\int \big[ \tilde{\br}' \in \bR A\big] d\tilde{r}_y'\big)^{\frac{1}{2}}},
\end{align}
the left- and right-hand sides of \eqref{spectralDecomposition} become respectively \eqref{changeOfVariableViaRotation} and
\begin{align}
    \sum_{\ell} \lambda_\ell \tilde{u}_\ell(\tilde{\br}') \overline{\tilde{u}_\ell(\tilde{\br})}, \label{inFactSpectralDecomposition}
\end{align}
where
\begin{align}
    \tilde{u}_\ell(\tilde{\br}) = \frac{\big[\tilde{\br} \in \bR A\big]}{\big(\int \big[\tilde{\br} \in \bR A\big] d\tilde{r}_y\big)^{\frac{1}{2}}} u_\ell(\tilde{r}_x). \label{orthonormalFunctions}
\end{align}
We can easily verify the orthonormality of \eqref{orthonormalFunctions}
from 
\begin{align}
    &\iint \frac{\big[\tilde{\br} \in \bR A\big]}{\int \big[\tilde{\br} \in \bR A\big] d\tilde{r}_y} u_\ell(\tilde{r}_x)  \overline{u_{\ell'}(\tilde{r}_x)}
    d\tilde{r}_x d\tilde{r}_y \label{twoDimensionSpectralStart}\\
    &=\int \bigg(\int \frac{\big[\tilde{\br} \in \bR A\big]}{\int \big[\tilde{\br} \in \bR A\big] d\tilde{r}_y} d\tilde{r}_y\bigg) u_\ell(\tilde{r}_x)  \overline{u_{\ell'}(\tilde{r}_x)}
    d\tilde{r}_x   \\
    &= \int u_\ell(\tilde{r}_x)  \overline{u_{\ell'}(\tilde{r}_x)} d\tilde{r}_x,\label{twoDimensionSpectralEnd}
\end{align}
implying that \eqref{inFactSpectralDecomposition} is a spectral decomposition of $\cB_{\balpha}$. Thus, $\{\lambda_\ell\}$, the eigenvalues of $\tilde{\cB}_{\balpha}$, are also the eigenvalues of $\cB_{\balpha}$.

\section{}
\label{minMaxProof}
Consider a general statement. Let $\cL$ be a positive semi-definite Hilbert-Schmidt operator with kernel $L(t,t')$ and let $\cG$ be an operator satisfying $(\cG s)(t) = g(t)s(t)$ with $g \in L^2(\bbR)$ and $0\leq |g(t)| \leq 1$ for all $t$.
We wish to prove that
\begin{align}
    \lambda_\ell(\cG\cL\cG) \leq \lambda_n(\cL).
\end{align}
The min-max theorem for matrices \cite[Thm. 4.2.6]{horn2012matrix}, which can be naturally generalized to self-adjoint Hilbert-Schmidt
operators, does the trick. Applying it, it can be seen that
\begin{align}
    \lambda_\ell(\cG\cL\cG) &= \min_{\varphi \in \vspan(\psi_0, \ldots, \psi_\ell), \varphi \neq 0}  \frac{\langle \cG\cL\cG\varphi, \varphi \rangle}{\langle \varphi,\varphi\rangle}\\
    &= \min_{\varphi \in \vspan(\psi_0, \ldots, \psi_\ell), \varphi \neq 0}  \frac{\langle \cL\cG\varphi, \cG\varphi \rangle}{\langle \varphi,\varphi\rangle} \\
    &\leq \min_{\tilde{\varphi} \in \vspan(\cG\psi_0, \ldots, \cG\psi_\ell), \tilde{\varphi}\neq 0}  \frac{\langle \cL\tilde{\varphi}, \tilde{\varphi} \rangle}{\langle \tilde{\varphi},\tilde{\varphi}\rangle}\\
    &\leq \max_{\dim(S)=\ell+1} \min_{\tilde{\varphi} \in S, \tilde{\varphi}\neq 0}  \frac{\langle \cL\tilde{\varphi}, \tilde{\varphi} \rangle}{\langle \tilde{\varphi},\tilde{\varphi}\rangle}\\
    &= \lambda_\ell(\cL),
\end{align}
where $\psi_k$ is the eigenvector of $\cG\cL\cG$ corresponding to the $k$th largest eigenvalue. The first inequality follows from the substitution $\tilde{\varphi}=\cG\varphi$ and $\|\tilde{\varphi}\| \leq  \|\varphi\|$.

\section{}
\label{upaSquintFreeProof}

This appendix proves that \eqref{upaSquintFree} holds if and only if $p\geq 1$.

\subsection{Sufficiency}

The sufficiency %of $\Nrf$ to remove the beam squint
requires \eqref{upaSquintFree} to vanish if $p \geq 1$. Using \eqref{ulaSquintFree} and \eqref{sandwich}, 
\begin{align}
    \sum_{\ell\geq \lceil p \alpha^{\sf up}\rceil} \!\!\! \lambda_\ell(\cB_{\balpha}) & \leq L_3\sum_{\ell\geq \lceil p \alpha^{\sf up}\rceil}\lambda_\ell(\cB_{\alpha^{\sf up}})\\
    &=L_3 \alpha^{\sf up}(1-p)^+ + o(r).
\end{align}
Therefore,
\begin{align}
    \frac{\sum_{\ell\geq \lceil p \alpha^{\sf up}\rceil} \lambda_\ell(\cB_{\balpha})}{\sum_\ell \lambda_\ell(\cB_{\balpha})} 
    &\leq \frac{L_3 \alpha^{\sf up}(1-p)^+ + o(r)}{\|\balpha\|}
\end{align}
which does converge to zero if $p\geq 1$. Recall that $\frac{\alpha^{\sf up}}{\|\balpha\|}$ is constant with respect to $r$.

\subsection{Necessity}
The necessity requires \eqref{upaSquintFree} not to vanish for $p < 1$. From \eqref{ulaSquintFree} and \eqref{sandwich}, %we similarly have
\begin{align}
    &\frac{\sum_{\ell\geq \lceil p \alpha^{\sf up}\rceil} \lambda_\ell(\cB_{\balpha})}{\sum_\ell \lambda_\ell(\cB_{\balpha})} \geq \frac{\delta L_3\alpha^{\sf lo} (1-\frac{\alpha^{\sf up}}{\alpha^{\sf lo}}p)^+ + o(r)}{\|\balpha\|}. \label{tailLower}
\end{align}
Recall that both $\frac{\alpha^{\sf lo}}{\|\balpha\|}$ and $\frac{\alpha^{\sf up}}{\alpha^{\sf lo}}$ are constant with respect to $r$.
From $p<1$, a small enough $\delta>0$ can be chosen such that
\begin{align}
    \frac{\alpha^{\sf up}}{\alpha^{\sf lo}}p = \frac{L_1}{(1-\delta)L_1+\delta L_2}p <1.
\end{align}
Therefore, \eqref{tailLower} cannot vanish.

\section{}
\label{partiallyConnectedProperty}

One welcome property of a partially-connected architecture is that, if the subarrays are identical, both in topology and in the number of phase shifts, one can replace their individual RF chains with delay lines and connect all of them to a single RF chain without loss in performance. While the result for linear arrays with MRT beamformer can be found in \cite[Sec. I-B]{mailloux1982phased} (also in \cite[Prop. 1]{dovelos2021channel} and \cite[Sec. III-D]{dai2022delay}), this appendix extends the result for completeness.

Precisely, the first condition corresponds to
\begin{align}
    \ba_m^*(f) = \exp \! \bigg(j2\pi \frac{\fc+f }{c}\bu^\top \overline{\br}_m\bigg) \ba_0^*(f) \label{subarrayChannel}
\end{align}
where $\overline{\br}_m$ is the displacement of the $m$th subarray with respect to the $0$th one (recall \eqref{losChannel}).
The second condition is
\begin{align}
    \bW_{\sf a} = \blkdiag(\bw_{\sf a},\ldots,\bw_{\sf a}) \label{blockDiagonalIdentical}
\end{align}
where $\bw_{\sf a}\in \bbC^{\frac{N}{M}}$ is the subarray beamformer.
%\angel{Not clear what "being equal in phase shifters" means.} \heedong{Clarified but still unsatisfactory.}

From the block diagonal structure \eqref{blockDiagonalIdentical}, the beamforming gain in \eqref{gainDefinition} is reduced to
\begin{align}
    g(f) = \frac{\big|\big(\sum_m \ba_m^*(f)[\bw_{\sf d}(f)]_m\big)\bw_{\sf a} \big|^2}{\|\bw_{\sf a}\|^2 \|\bw_{\sf d}(f)\|^2},
\end{align}
Armed with \eqref{subarrayChannel}, it can be recast as
\begin{align}
    \frac{\big|\sum_m \exp \! \big(j2\pi \frac{\fc+f }{c}\bu^\top \overline{\br}_m\big)[\bw_{\sf d}(f)]_m \big|^2}{\|\bw_{\sf d}(f)\|^2} \cdot \frac{|\ba_0^*(f)\bw_{\sf a}|^2}{\|\bw_{\sf a}\|^2}.
\end{align}
Applying the Cauchy-Schwarz inequality,
\begin{align}
    g(f) \leq Mg_0(f),
\end{align}
and the equality is seen to be attained by 
\begin{align}
    [\bw_{\sf d}(f)]_m = \exp \! \bigg(-j2\pi \frac{\fc+f }{c}\bu^\top \overline{\br}_m\bigg), \label{delayLines}
\end{align}
which can be implemented with delay lines. This maximum beamforming gain 
equals \eqref{averageGainHybridlyConnected}, obtained with $M$ RF chains.

The resulting architecture with delay lines mitigates the beam squint across the subarrays. The remainder, the beam squint within the subarray, is relatively negligible.
% \cite[Sec. 8.3.2]{mailloux2005phased}.
This very idea has recently applied to a multipath setting in \cite{dovelos2021channel,dai2022delay}.

\bibliographystyle{IEEEtran}
\bibliography{jour_short,conf_short,ref}

\end{document}

%% file: input.tex
\def\Nt{{N_{\mathrm{t}}}}
\def\Nr{{N_{\mathrm{r}}}}
\def\Nrf{{N_{\sf RF}}}
\def\fc{{f_{\mathrm{c}}}}
\def\gavg{{g_{\sf avg}}}

\def\vspan{{\sf span}}
\def\blkdiag   {\mbox{\rm blkdiag}}

\def\bb0{{\mathbb{0}}}

\def\ba{{\boldsymbol{a}}}
\def\bb{{\boldsymbol{b}}}

\def\br{{\boldsymbol{r}}}

\def\bu{{\boldsymbol{u}}}
\def\bv{{\boldsymbol{v}}}
\def\bw{{\boldsymbol{w}}}

\def\b0{{\boldsymbol{0}}}

\def\bB{{\boldsymbol{B}}}

\def\bH{{\boldsymbol{H}}}

\def\bR{{\boldsymbol{R}}}

\def\bW{{\boldsymbol{W}}}

\def\b{{\mathrm{b}}}

\def\r0{{\mathbf{0}}}

\def\bbC{{\mathbb{C}}}

\def\bbR{{\mathbb{R}}}

\def\cB{\mathcal{B}}

\def\cG{\mathcal{G}}

\def\cL{\mathcal{L}}

\def\cO{\mathcal{O}}

\def\balpha{\bm \alpha}

\def\bsf0{{\bm{\mathsf{0}}}}

\def\Pt{{P_{\mathrm{t}}}}

\def\N0{{N_{\mathrm{0}}}}

\def\kron{\otimes}

\def\fc{f_{\mathrm{c}}}

\def\bsf{{\boldsymbol{s}_\mathrm{f}}}

\newcommand{\be}{\begin{equation}}
\newcommand{\ee}{\end{equation}}
\newcommand{\bal}{\begin{align}}
\newcommand{\eal}{\end{align}}
\def\tr {{\rm tr}}

\def\SNR    {{\mathsf{SNR}}}

\def\Gt {G_{\mathrm{t}}}
\def\Gr {G_{\mathrm{r}}}

\def\fc {f_{\mathrm{c}}}